\pdfoutput=1
\documentclass[11pt,letterpaper]{scrartcl}

\makeatletter
\DeclareOldFontCommand{\rm}{\normalfont\rmfamily}{\mathrm}
\DeclareOldFontCommand{\sf}{\normalfont\sffamily}{\mathsf}
\DeclareOldFontCommand{\tt}{\normalfont\ttfamily}{\mathtt}
\DeclareOldFontCommand{\bf}{\normalfont\bfseries}{\mathbf}
\DeclareOldFontCommand{\it}{\normalfont\itshape}{\mathit}
\DeclareOldFontCommand{\sl}{\normalfont\slshape}{\@nomath\sl}
\DeclareOldFontCommand{\sc}{\normalfont\scshape}{\@nomath\sc}
\makeatother

\usepackage[top=2.5cm, bottom=2.5cm, left=3.2cm, right=3cm]{geometry}

\usepackage[T1]{fontenc}
\usepackage[utf8]{inputenc}

\usepackage[hidelinks]{hyperref}
\usepackage[parfill]{parskip}
\usepackage[separate-uncertainty = true,multi-part-units=single]{siunitx}
\usepackage{slashed}
\usepackage{microtype}
\usepackage{csquotes}
\usepackage{graphicx}
\usepackage{amsmath}
\usepackage{amssymb}
\usepackage{braket}
\usepackage{booktabs}
\usepackage{subcaption}
\usepackage{siunitx}
\usepackage{todonotes}
\usepackage{nicefrac}

\usepackage[numbers,sort&compress]{natbib}

\usepackage{multirow}

\usepackage{fancyhdr}
\usepackage[yyyymmdd,hhmmss]{datetime}
\fancypagestyle{plain}
{
}
\usepackage{cleveref}

\usepackage[auth-sc,affil-it]{authblk}

\usepackage{scalefnt}
\newcommand{\abbrev}{\scalefont{.9}}

\newcommand{\NNLO}{\text{\abbrev NNLO}}
\newcommand{\NLO}{\text{\abbrev NLO}}
\newcommand{\LO}{\text{\abbrev LO}}

\newcommand{\SMEFT}{\text{\abbrev SMEFT}}

\newcommand{\SM}{\text{\abbrev SM}}

\newcommand{\IR}{\text{\abbrev IR}}

\newcommand{\QCD}{\text{\abbrev QCD}}

\newcommand{\CKM}{\text{\abbrev CKM}}
\newcommand{\PDF}{\text{\abbrev PDF}}

\newcommand{\LHC}{\text{\abbrev LHC}}

\newcommand{\CTFOURTEEN}{\text{\abbrev CT14}}

\newcommand{\DIS}{\text{\abbrev DIS}}

\newcommand{\MCFM}{\text{\abbrev MCFM}}
\newcommand{\DDIS}{\text{\abbrev DDIS}}

\newcommand{\BCM}{\text{\abbrev BCM}}
\newcommand{\BGZ}{\text{\abbrev BGZ}}
\newcommand{\BCST}{\text{\abbrev BCST}}
\newcommand{\GLZ}{\text{\abbrev GLZ}}

\newcommand{\abs}[1]{\lvert#1\rvert}

\newcommand{\taucut}{\ensuremath{\tau^\text{cut}}}
\newcommand{\taucuth}{\ensuremath{\tau_h^\text{cut}}}
\newcommand{\taucutd}{\ensuremath{\tau_d^\text{cut}}}
\newcommand{\taucutl}{\ensuremath{\tau_l^\text{cut}}}

\newcommand{\cosxprod}{\ensuremath{\cos \theta_{l,x}}}
\newcommand{\cosyprod}{\ensuremath{\cos \theta_{l,y}}}
\newcommand{\coszprod}{\ensuremath{\cos \theta_{l,z}}}

\newcommand{\coslstar}{\ensuremath{\cos \theta_{l}^*}}
\newcommand{\coslN}{\ensuremath{\cos \theta_{l}^N}}
\newcommand{\coslT}{\ensuremath{\cos \theta_{l}^T}}

\newcommand{\cosphiT}{\ensuremath{\cos\phi^T}}
\newcommand{\cosphiN}{\ensuremath{\cos\phi^N}}

\newcounter{notecount}

\makeatletter
\renewcommand\maketitle{
	\begin{center}
		{\huge\bfseries\@title\par\vspace{0.5em}}
		{\scshape\@author}
	\end{center}
}
\makeatother

\fancypagestyle{firstpage}{%
	\lhead{}
	\rhead{FERMILAB-PUB-20-608-T, IIT-CAPP-20-05}
}

\begin{document}

\thispagestyle{firstpage}
\title{\LARGE Single-top-quark production in the $t$-channel at \NNLO{} }

\author[1]{John Campbell}
\author[1,2,3]{Tobias Neumann}
\author[2]{Zack Sullivan}

\affil[1]{Fermilab, PO Box 500, Batavia, Illinois 60510, USA}
\affil[2]{Department of Physics, Illinois Institute of Technology, Chicago, Illinois 60616, USA}
\affil[3]{Department of Physics, Brookhaven National Laboratory, Upton, New York 11973, USA}

\maketitle

\vspace{1cm}

\begin{abstract}
We present a calculation of $t$-channel single-top-quark production and decay in 
the five-flavor scheme at \NNLO{}. Our
results resolve a disagreement between two previous calculations of this process that
found a difference in the inclusive cross section at the level of the \NNLO{}
coefficient itself. We compare in detail
with the previous calculations at the inclusive, differential and fiducial level including
$b$-quark tagging at a fixed scale $\mu=m_t$.  In addition, we advocate
the use of double deep inelastic scattering (\DDIS{}) scales ($\mu^2=Q^2$ for
the light-quark line and $\mu^2=Q^2+m_t^2$ for the heavy-quark line)
that maximize perturbative stability and allow for robust scale uncertainties.   
All \NNLO{} and $\NLO{}\otimes\NLO{}$ contributions for production and decay are 
included in the on-shell and vertex-function approximation. 
We present fiducial and differential results for a
variety of observables used in Standard Model and Beyond Standard Model analyses, 
and find an important difference between the \NLO{} and \NNLO{} predictions of exclusive 
$t+n$-jet cross sections. Overall we find that \NNLO{} corrections are crucial for a precise 
identification of the $t$-channel process.
\end{abstract}

\tableofcontents
\clearpage

\fancyhead[LE,RO]{\textsl{\rightmark}}
\fancyhead[LO,RE]{}

\section{Introduction}
\label{sec:introduction}

Top quarks play a special role in the Standard Model (\SM{}).  They stand out from the other
quarks by virtue of their mass, significantly larger than their peers, and by the fact
that they decay before they are able to form bound states.  The former property ensures
that top quarks are primarily responsible for the production of Higgs bosons at hadron colliders,
by mediating the loop-induced coupling $gg \to h$, and positions them as potential portals
to so-far undiscovered extensions of the \SM{}.  The latter enables both a clean
theoretical description of top-quark processes and, on the experimental side, their
identification through decays into well-measured objects.
Therefore the study of the production and decay of top quarks is a cornerstone of the current
and future physics program of the Large Hadron Collider (\LHC{}). 

The primary mechanism for producing top quarks is through the strong interaction, resulting in
the creation of a top quark-antiquark pair.  However, a single top quark may also be
produced through a weak interaction involving a bottom quark.  In fact, the $t$-channel 
single top production processes, represented at leading order by,
\begin{equation}
q + \overset{(-)}{b} \to q^\prime + \overset{(-)}{t} \,,
\label{eq:nodecay}
\end{equation}
and first observed at the Tevatron~\cite{Abazov:2009ii,Aaltonen:2009jj}, are responsible
for approximately $20\%$ of all top-quark production at the \LHC{}.  Despite
this production mode proceeding through weak couplings, the rate is large
due to its $t$-channel nature and the fact that it is kinematically favored
compared to the pair-production process.  The channel of course provides a useful probe
of the top quark itself, with measurements of the top-quark mass~\cite{Sirunyan:2017huu}
and polarization~\cite{Khachatryan:2015dzz}, detailed tests of the Standard Model at the
differential level~\cite{Aaboud:2017pdi,Sirunyan:2019hqb}, as well as constraints on
anomalous $Wtb$ couplings~\cite{Aad:2015yem,Khachatryan:2016sib,Aaboud:2017aqp}.
In addition, it can also provide valuable information on the elements of the
production mechanism: the bottom quark parton distribution function (\PDF{}) and the \CKM{} matrix 
element, $V_{tb}$.  Indeed, measurements of $V_{tb}$ have been made at both the
Tevatron~\cite{Aaltonen:2015cra} and the
LHC~\cite{Chatrchyan:2012ep,Chatrchyan:2011vp,Khachatryan:2014iya,Sirunyan:2020xoq}.

In order to turn experimental observations into precision measurements it is essential to
have theoretical calculations with small residual uncertainties.  One of the largest sources
of theoretical uncertainty results from the truncation of the perturbative expansion at
a fixed order and is usually estimated by scale variation.  Generically, this scale variation
is expected to decrease order-by-order, with percent-level uncertainties only expected
when going to the current gold-standard for precision measurements,
next-to-next-to-leading order (\NNLO{}).  For top-quark processes another consideration
that complicates the perturbative calculation stems from the fact that the top quark decays,
$t \to W^+ b$.  Accounting for such a decay, at leading order, does not present a tremendous
complication -- in fact, using the spinor helicity approach, expressions for relevant
matrix elements are extremely compact~\cite{Kleiss:1988xr}.  However, once strong-coupling
corrections are included one should, in principle, account for the effects of virtual radiation
that connects strongly-interacting particles in the `production' and `decay' elements
of the process.

The earliest results for $t$-channel single top production that included corrections to
next-to-leading order (\NLO{}) in the strong coupling were computed for a stable top quark, that is,
for exactly the processes shown in 
\cref{eq:nodecay}~\cite{Bordes:1994ki,Stelzer:1997ns,Kant:2014oha}.
These were soon extended beyond the case of the inclusive cross section, to also describe
fully-differential measurements of this 
process~\cite{Harris:2002md,Sullivan:2004ie,Sullivan:2005ar}.
Somewhat later the decay of the top quark was also included at the same order, with
calculations performed in a factorized approach in which the top quark remains on its mass shell
and radiation that connects production and decay is
neglected~\cite{Campbell:2004ch,Cao:2004ky,Schwienhorst:2010je}.  Although this is an inherent 
approximation,
for sufficiently inclusive observables such off-shell effects are small, of the order of
$\Gamma_t/m_t$~\cite{Fadin:1993kt,Melnikov:1993np}.   For a precision prediction of the invariant 
mass of the top quark it
is clearly essential to move away from this approximation and, in general, more differentially such 
effects
are of increasing importance.  Calculations of off-shell effects were first performed in an 
effective
field theory approach, valid in the region of the
resonance~\cite{Pittau:1996rp,Beneke:2004km,Falgari:2011qa,Falgari:2010sf},
and subsequently computed exactly~\cite{Papanastasiou:2013dta,Frederix:2016rdc,Neumann:2019kvk}.

The first steps towards including the effects of \NNLO{} corrections to this process were taken in
refs.~\cite{Gao:2012ja,Brucherseifer:2013iv}, in which the top-quark decay was computed at this
order.  The corresponding \NNLO{} corrections for the hadroproduction of a stable top quark 
in this channel were subsequently computed~\cite{Brucherseifer:2014ama} and, finally,
the full corrections in both the production and decay stages were included in the
calculation presented in refs.~\cite{Berger:2016oht,Berger:2017zof}.  The results of these
calculations indicate that, at the inclusive level, the effect of the \NNLO{} corrections is
small, leading to a change in the cross section of only a few percent.  However, the effects
of the corrections can be larger in the fiducial volume~\cite{Berger:2016oht,Berger:2017zof}.
Importantly, a comparison between the two groups -- at the level of inclusive cross sections for
a stable top quark -- revealed that the two calculations disagree at the 1\% 
level~\cite{Berger:2016oht}, that is, at almost the same level as the effect of the \NNLO{} 
corrections themselves. Recently, the \NNLO{} calculation of 
refs.~\cite{Berger:2016oht,Berger:2017zof} has 
been used to study the top-quark mass 
determination \cite{Yuan:2020nzd} and differences between flavor schemes \cite{Gao:2020ejr}. 

The importance of this production mode demands that the existing theoretical calculations be
scrutinized, cross-checked and any discrepancies understood.  To this end, in this paper we present
a full re-calculation of the \NNLO{} \QCD{} corrections to this process, including all the effects of
radiation in top-quark production and decay at this order.  Although the calculation necessarily
shares some elements and methodology with previous work, it has been performed from scratch from
ingredients that have been, where possible, verified against independent computations. 
In order to ensure the maximum value
of the calculation, it is performed using the framework of the \MCFM{}
package~\cite{Campbell:2010ff,Campbell:2015qma,Campbell:2019dru} and will be distributed publicly.
The aim of this approach is to ensure that the latest possible theoretical information can always
be incorporated in future experimental analyses, incorporating any demands on input parameters
such as the top-quark mass and \PDF{}s.\footnote{For a recent example where theoretical input was
limited to the NLO level, justified by the lack of such flexibility, see 
ref.~\cite{Sirunyan:2019hqb}.}

The outline of the paper is as follows. In \cref{sec:calculation} we describe the setup of our 
calculation: outlining the general structure, discussing the role of a 
strict fixed-order expansion that becomes relevant when including the top-quark decay and
introducing 
the distinctive choice of 
factorization and renormalization scales for this process. We also describe our prescription for 
$b$-quark tagging, which at \NNLO{} needs special attention. We then present details of
all the necessary ingredients for the three independent \NNLO{} sub-calculations entering our 
results, as well as of the $\NLO{}\otimes\NLO{}$ interference contributions, and describe
the checks that we have performed to validate our results. In 
\cref{sec:comparison} we 
compare in-depth with the previous stable-top calculation and the on-shell calculation with decay.
Last, in \cref{sec:results} we discuss fully inclusive cross sections, cross sections with fiducial 
cuts, and the relevant differential distributions used in experimental analyses. We 
furthermore consider \NNLO{} results 
for angular observables in the top-quark rest frame, angles that are used for searches of 
physics 
beyond the Standard Model and which are sensitive to $Wtb$-vertex modifications in the production and 
decay stages.

\section{Calculation}
\label{sec:calculation}

\begin{figure}
	\centering
	\includegraphics[width=2.5in]{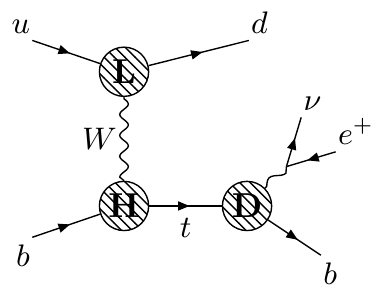}
	\caption{Feynman diagram for $t$-channel production in the on-shell approximation 
	with blobs denoting vertex corrections on the light line (L), on the heavy line in production 
	(H) and in decay (D). }
	\label{fig:process}
\end{figure}

\begin{figure}
	\centering
	\includegraphics[width=2.5in]{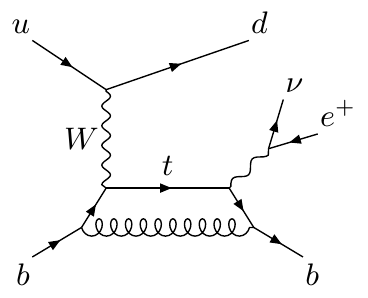}
	\includegraphics[width=2.5in]{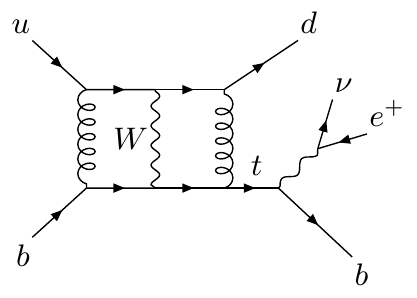}
	\caption{Example Feynman diagrams for one-loop off-shell contributions (left) and 
	color-suppressed 
	interference contributions (right) in $t$-channel production.}
	\label{fig:process2}
\end{figure}

Our calculation of \NNLO{} \QCD{} corrections to $t$-channel single-top-quark production and decay is 
performed in the 
five-flavor scheme, a $2\to4$ scattering process at Born level that is depicted schematically
in \cref{fig:process}.   In order to render the calculation tractable
we perform the calculation in the 
on-shell 
approximation, neglecting contributions from radiation that connects the production
and decay stages such as the diagram shown in \cref{fig:process2} (left).  For a large class of
observables this amounts to discarding off-shell effects that are of order
$\Gamma_t/m_t$ \cite{Fadin:1993kt,Melnikov:1993np}. We 
furthermore neglect a class
of two-loop box diagrams that connect the light-quark and heavy-quark lines, that are suppressed by
a color factor
of $1/N_c^2$, see \cref{fig:process2} (right).\footnote{While there has been progress on the reduction of 
these two-loop integrals to a basis of master integrals \cite{Assadsolimani:2014oga}, the solution 
of the integrals themselves is still an open issue.} The \NNLO{} corrections therefore consist of 
\NNLO{} vertex corrections on the 
light-quark line and the heavy-quark line in the production part of the process, as well as 
\NNLO{} 
vertex corrections in the decay of the top quark, as indicate in \cref{fig:process}. In the literature this is also known as 
the structure function approximation. Of 
course, in addition to second-order corrections to each vertex, at \NNLO{} we must also include
one-loop times one-loop interference contributions.

We calculate each of the three \NNLO{} vertex corrections in \cref{fig:process} using slicing-based 
subtractions derived from 
different factorization theorems. The interference contributions are calculated in a mixed 
slicing-dipole subtraction scheme.
While we use the same \NNLO{} factorization theorem to assemble the corrections on the heavy line 
in production and decay as in a previous calculation \cite{Berger:2016oht,Berger:2017zof}, we have 
performed a series of exhaustive analytical and numerical checks to ensure a correct cancellation 
of \IR{} divergences between all amplitudes. Our implementation of each contribution is,
of course, completely independent and all amplitudes have either been calculated
from scratch or obtained from  different sources.
With this we can guarantee a fully independent cross-check of previous results. 
The calculational details for the three \NNLO{} vertex corrections and the $\NLO{}\otimes\NLO{}$ 
interference are given in the next four subsections, but we first discuss some general aspects of 
the calculation.

\paragraph{Fixed-order expansion.}

Since the top quark can be treated as a quasi-stable particle to a good approximation, the 
differential cross section $\mathrm{d}\sigma(\alpha_s)$ can be factorized into a production part 
$\mathrm{d}\sigma^\text{production}(\alpha_s)$ and a decay part $ 
\mathrm{d}\Gamma_t(\alpha_s)/\Gamma_t(\alpha_s)$. In the decay corrections the numerator is the 
differential decay 
width and the denominator is its integrated expression, such that the integration of the
decay fully inclusively leads to no change in the cross section.
Throughout this paper we assume that the top-quark decays 100\% of the time to a $W$ boson and a 
$b$~quark.

While it is possible to keep the full expression of the width in the denominator, it is customary 
to expand the whole cross section $\mathrm{d}\sigma(\alpha_s)$ in $\alpha_s$ to 
avoid higher-order contributions from the expansion of the width in the denominator and the 
factorized contributions of the individual vertex corrections. Consequently, we expand the fully 
differential cross section $\mathrm{d}\sigma(\alpha_s) = 
\mathrm{d}\sigma^\text{production}(\alpha_s) \otimes \mathrm{d}\Gamma_t(\alpha_s) / 
\Gamma_t(\alpha_s)$ 
consistently in $\alpha_s$ to obtain cross sections at \LO{} ($\alpha_s^0$), \NLO{} ($\alpha_s$) 
and \NNLO{} ($\alpha_s^2$). The symbol $\otimes$ denotes that production and decay are taken fully
at the amplitude level including all spin correlations. Denoting with $\mathrm{d}\sigma^{(j)}$ the 
contribution
of order $\alpha_s^j$ in production and with $\mathrm{d}\Gamma_t^{(k)}$ the contribution of order 
$\alpha_s^k$ in decay, the fixed-order expansion is explicitly given by:
\begin{align}
	\begin{split}
	\mathrm{d}\sigma_\LO{}  &  = \frac{1}{\Gamma_t^{(0)}} 
	\mathrm{d}\sigma^{(0)}\otimes\mathrm{d}\Gamma_t^{(0)} \,,\\
	\mathrm{d}\sigma_{\delta\NLO{}} & = \frac{1}{\Gamma_t^{(0)}} \Bigg[ 
			\mathrm{d}\sigma^{(1)}\otimes\mathrm{d}\Gamma_t^{(0)} + 
			\mathrm{d}\sigma^{(0)}\otimes\left(\mathrm{d}\Gamma_t^{(1)} 
			- \frac{\Gamma_t^{(1)}}{\Gamma_t^{(0)}} \mathrm{d}\Gamma_t^{(0)}  \right)
			\Bigg] \,,\\
	\mathrm{d}\sigma_{\delta\NNLO{}} & = \frac{1}{\Gamma_t^{(0)}} \Bigg[
			\mathrm{d}\sigma^{(2)}\otimes\mathrm{d}\Gamma_t^{(0)} + 
			\mathrm{d}\sigma^{(1)}\otimes\left(
					\mathrm{d}\Gamma_t^{(1)} - 
					\frac{\Gamma_t^{(1)}}{\Gamma_t^{(0)}}\mathrm{d}\Gamma_t^{(0)}
				\right) \\
		& \qquad\qquad	+ \mathrm{d}\sigma^{(0)}\otimes \left(
					\mathrm{d}\Gamma_t^{(2)} - 
					\frac{\Gamma_t^{(2)}}{\Gamma_t^{(0)}}\mathrm{d}\Gamma_t^{(0)}
					-\frac{\Gamma_t^{(1)}}{\Gamma_t^{(0)}} \left(
							\mathrm{d}\Gamma_t^{(1)} - \frac{\Gamma_t^{(1)}}{\Gamma_t^{(0)}}
								\mathrm{d}\Gamma_t^{(0)}
						\right)
				\right) \Bigg] \,,
	\label{eq:fixedorderexpansion}
	\end{split}
\end{align}
where $\Gamma_t^{(l)}$ are the integrated decay corrections of order $\alpha_s^k$.
By construction, the expressions in parentheses vanish upon fully inclusive integration
over the decay

\paragraph{Double deep inelastic scattering (\DDIS{}) scales.}

Our implementation allows for different factorization and renormalization scales in 
light-line corrections, heavy-line corrections in production, and corrections in decay. For our 
approximation in terms of vertex corrections the natural choice to minimize scale-dependent 
logarithms 
for fully inclusive results is 
to take $Q^2$ as the central scale
on the light-quark line, $Q^2+m_t^2$ on the heavy-quark line in production, and $m_t$ in the decay, 
where $Q^2$ is the 
positive squared momentum of the $W$ boson.
We abbreviate this set of scale choices as \DDIS{} (double deep inelastic scattering) scales.

The motivation for considering \DDIS{} scales stems from the observation that single-top-quark
production in our approximation is double deep inelastic scattering with a light and heavy quark.
When \PDF{}s are extracted from \DIS{} measurements, one therefore expects our perturbative results
to be stable across orders with \DDIS{} scales. That is, strictly speaking, the cross sections at 
\LO{}, \NLO{}
and \NNLO{} should agree within \PDF{} uncertainties. Therefore, at least in principle, using 
\DDIS{} scales can be exploited as a constraint for \PDF{}s. At \NNLO{}, off-shell effects and 
color-suppressed contributions from two-loop box diagrams break 
this property, see \cref{fig:process2}, but such effects are estimated to be small, at least 
inclusively.

\paragraph{$b$-quark tagging.}

Tagging a specific quark flavor in jets, say a $b$ quark, raises the question of 
infrared safety. Throughout \NLO{} there are no issues, but at \NNLO{} 
a large-angle soft gluon splitting $ g\to b\bar{b}$ can lead to $b$-quarks 
being 
clustered into different jets, violating flavor infrared safety. Typical jet algorithms only 
ensure infrared safety at the event momentum level, and this situation breaks infrared safety at 
the jet flavor level. A general solution
is to employ a flavor-jet algorithm for infrared-safe predictions \cite{Banfi:2006hf,Banfi:2007gu},
which ensures that such a soft splitting would recombine to a flavorless jet.

Experimentally such a flavor-jet algorithm is not used in single-top-quark analyses and we do not
adopt such an algorithm in this study.   Although it could, in principle, be interesting to consider 
its impact in the future, the difference is likely to be negligible due to the smallness of the 
$g\to b\bar{b}$ contributions, see below. For now we adopt the 
strategy used in ref.~\cite{Berger:2017zof}, which
also enables a more transparent comparison with the results presented in that reference.
The procedure is as follows: for the \NNLO{} vertex corrections in production the $g\to b\bar{b}$ 
splitting does not come with another $b$ quark at the same vertex. The flavor in the pair can 
therefore be ignored on 
the premise that it gives a tiny contribution. In the decay the situation with an additional $b$ 
quark from the top-quark decay arises and
the flavor is ignored for the $b\bar{b}$ pair with the smaller invariant mass, which is most likely 
the pair originating from the $g\to b \bar{b}$ splitting.

The authors of 
ref.~\cite{Berger:2017zof} check the infrared safety of this approach numerically by evaluating 
the infrared subtraction slicing-cutoff dependence. They find that when no flavors are ignored 
there is indeed a 
difference, but it is tiny. We do confirm these findings, and note that the minuscule size of this
effect, unless one probes tiny slicing cutoffs to enhance it, confirms that the flavor contribution 
of the $g\to b\bar{b}$ contribution is negligible. Therefore, using a full flavor-jet algorithm
does not seem necessary in practice.

We furthermore adopt the following choice of the $b$-jet definition for the remainder of this 
paper. Namely, we 
define the $b$-number of a jet to be the sum of the $b$-numbers of its constituent partons,
where the $b$-number of a $b$ and $\bar b$ quark are $+1$ and $-1$, respectively.  A jet with
non-zero $b$-number is termed a $b$-jet and all other jets are light jets.

\paragraph{Common ingredients.}
Before detailing the individual \NNLO{} calculations, we first summarize some ingredients shared by 
all parts. For the calculation of 
the \NNLO{} vertex corrections, the above-cut slicing contributions are obtained by performing 
\NLO{} 
calculations of partonic processes corresponding to Born configurations plus an additional
parton. Importantly, these have to be numerically stable in the 
doubly-singular limits. 
The necessary 
tree-level and one-loop amplitudes for these calculations are assembled from existing amplitudes in 
\MCFM{} for 
one-loop corrections \cite{Campbell:2004ch} and tree-level results in the four-flavor scheme 
\cite{Campbell:2012uf}. We use dipole subtractions to combine them into \NLO{} calculations for
massless~\cite{Catani:1996vz} and massive~\cite{Catani:2002hc} particles.
The extension to limit the size of subtractions provides an additional check of a correct 
implementation (\enquote{alpha independence}) and can improve numerical stability. These extensions
have been worked out for both massless \cite{Nagy:1998bb,Nagy:2003tz} and massive 
\cite{Campbell:2004ch,Campbell:2005bb,Bevilacqua:2009zn} partons. Note that to retain a proper 
definition of the \DDIS{} scales, the dipoles in our \NLO{} calculations must be 
chosen with some care so that no initial-initial dipoles transform momenta on both lines.

\subsection{Top-quark-production corrections on the light line}

For the light-quark line, which represents \DIS{}-like jet production, we compute the \NNLO{} 
corrections using a factorization theorem in the 1-jettiness variable $\tau_l$ for massless partons 
\cite{Stewart:2010tn}. This approach has been first used in ref.~\cite{Boughezal:2015dva} for a 
calculation of $W$+jet at \NNLO{} and the formalism including all necessary ingredients at \NLO{} 
has been presented in full generality in ref.~\cite{Gaunt:2015pea}.

At \NNLO{} we have to take into account contributions with one and two extra emissions in addition 
to the parton already present at Born level. We denote with $M$ the number of (light-line) partonic 
emissions $p_k^\mu$, $k=1,\ldots,M$ and the initial state light-quark (beam) momentum as $q_b$. 
Then our 1-jettiness variable is defined as
\begin{equation}
	\tau_l = \frac{1}{Q_0} \sum_{k=1}^M \min\{ n_b p_k, n_j p_k \} \,
	\label{eq:taul}
\end{equation}
where $n_b$ is the normalized beam direction $q_b^\mu$ and $n_j$ is a normalized jet direction.
The scale $Q_0$ is introduced to make $\tau_l$ dimensionless, as discussed further below.
Obtaining the jet direction $n_j$ from an explicit minimization of $\tau_l$ for one and two 
additional
emissions has been discussed in refs.~\cite{Gaunt:2015pea,Stewart:2010tn}. For example for $M=2$
the result is
\begin{equation}
	\tau_l = \frac{1}{Q_0} \min \left\{ \min\{ E_1-p_1^z, E_2-p_2^z \},\, E_1+E_2 - 
	|\vec{p}_1+\vec{p}_2| 
	\right\} \,,
\label{eq:taulmin}
\end{equation}
effectively partitioning the phase space into extra emission clustered with the beam, or together
with the other final state emission. Here $p_k^z$ is the $z$-component of $p_k^\mu$,
$E_k$ is the energy of $p_k^\mu$, and $\vec{p}_k$ is the spatial component 
of $p_k^\mu$. The case of $M=3$ results
in a similar partitioning, where now all extra emissions can either be together, or with the beam,
or one with the beam and one with the other emission, in various combinations.  For further
details, see ref.~\cite{Gaunt:2015pea} or our implementation in \MCFM{}.

For the plots presented in this study we choose $Q_0=\SI{1}{\GeV}$. An alternative choice is
$Q_0\propto \sqrt{Q^2}$, where $Q^2$ is the 
squared 
$W$ boson 
momentum transfer. We find that at $\sqrt{s}= \SI{13}{\TeV}$ the choice of $Q_0 \sim 
\sqrt{Q^2}/100$ has similar numerical stability as $Q_0=\SI{1}{\GeV}$ and is subject to power 
corrections
that are of the same magnitude, as may be anticipated from the explicit minimization indicated in \cref{eq:taulmin}.
We do not explore this difference any further here.

Instead of the explicit minimization of $\tau_l$, in principle any infrared-safe Born projection 
can be used
to obtain the jet axis $n_j$. For example one could determine it from the jet clustering algorithm, 
as has previously been done in \MCFM{}. Any difference in the Born projection results
in different power-suppressed terms in the factorization theorem, but does not modify its singular 
structure \cite{Stewart:2010tn}.\footnote{In ref.~\cite{Boughezal:2019ggi} it has been 
found numerically for $Z$+jet production at \NLO{}
that the explicit minimization reduces power-suppressed corrections relative to a jet axis
obtained from the anti-$k_T$ jet algorithm.}

With this phase-space partitioning, one can write down the $1$-jettiness factorization theorem
integrated over $\tau_l$ as a subtraction scheme as
\begin{multline}
\mathrm{d}\sigma_{a,b}(\tau_l < \taucutl) = \mathrm{d}\sigma^\text{Born}\, f_b(x_b,\mu) 
\int_0^{\taucutl} 
\!\!\!\mathrm{d}\tau_l' 
\int\!\mathrm{d}k_j \int\!\mathrm{d}k_s \int\!\mathrm{d}t_a\, B_a(t_a,x_a,\mu)\, J(k_j,\mu) \times \\ 
S(k_s,\mu) \otimes H(\mu) \,
	\delta\left(\tau_l'- \frac{t_a}{E_a} - \frac{k_j}{E_j} - k_s \right) \,, \qquad\qquad
	\label{eq:jettiness_light}
\end{multline}
where we denote color correlations between hard function $H$ and soft function $S$ with $\otimes$. 
Parton flavors are labeled as $a$ and $b$ and the fully differential Born-level cross section 
$\mathrm{d}\sigma^\text{Born}$ has to be understood without \PDF{}s.

The beam functions $B_a$ have been calculated up to \NNLO{} 
in
ref.~\cite{Gaunt:2014xga} (for \NLO{} see refs.~\cite{Stewart:2009yx,Stewart:2010qs}) and 
previously implemented in \MCFM{} \cite{Boughezal:2016wmq}. The jet function $J$ has been 
calculated in 
ref.~\cite{Gaunt:2015pea}. The hard function $H$ corresponds to a crossed version of the \NNLO{} 
(two 
loop) $q\bar{q}\to V$ form factor \cite{Gehrmann:2005pd}. Lastly, we use the \NNLO{} soft function 
$S$ from ref.~\cite{Campbell:2017hsw} (calculated at \NLO{} in ref.~\cite{Jouttenus:2011wh}) as 
used in 
ref.~\cite{Campbell:2019gmd} for $H$+jet production, see also ref.~\cite{Boughezal:2015eha}. 
$f_b(x_b,\mu)$ are the usual \PDF{}s with parton momentum fraction $x_b$ of parton flavor $b$ for 
the $b$-quark line.

We perform the integrations and distributional convolutions over $\tau_l'$, $k_s$ and $k_j$ in 
\cref{eq:jettiness_light} explicitly in Laplace space and implement the result as the below-cut
contribution. The full \NNLO{} result is obtained by adding
the \NLO{} calculation with an additional jet in the presence of a small slicing cutoff $\tau_l$ 
and considering either 
the limit $\tau_l\to 0$ and/or a sufficiently small value of $\tau_l$ such that power-suppressed
corrections of $\mathcal{O}(\tau_l \log^k \tau_l)$ to the factorization theorem are negligible.

The matrix elements entering the calculation of the above-cut contribution
are easily computed.  We have recycled results from existing calculations
of $W+$jet production at \NLO{}~\cite{Bern:1997sc} that are implemented in
\MCFM{}, attaching a current $W \to t \bar b$,
in order to obtain the relevant amplitudes.
As a cross-check, the matrix elements
have been separately evaluated using the Recola package~\cite{Actis:2016mpe}
to ensure the correctness of all contributions.

\begin{figure}
	\centering
	\includegraphics{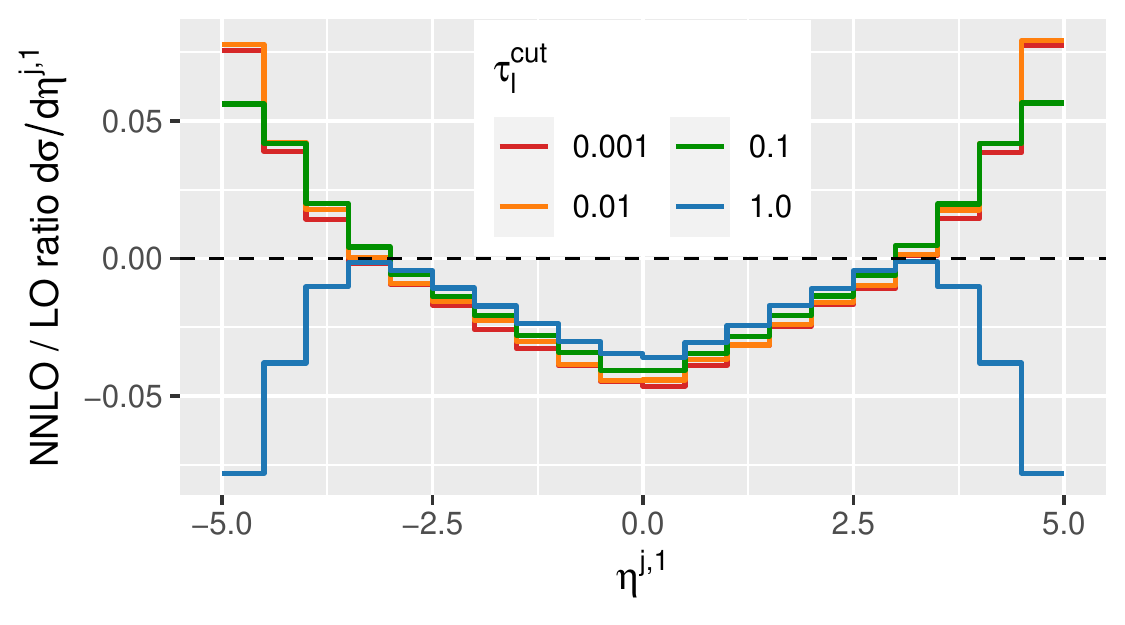}
	\caption{\NNLO{} corrections from the light-line production vertex for the leading jet 
		pseudorapidity distribution, relative to 
		\LO{}. The $\taucutl$-dependence of the corrections is shown in the range
		$0.001$--$1.0$.}
	\label{fig:lightetaj1}
\end{figure}

In \cref{fig:lightetaj1} we show the $\taucutl$-dependence of the \NNLO{} corrections for the 
leading jet pseudorapidity distribution as a fraction of the \LO{} distribution. We use the 
anti-$k_T$ jet clustering with $R=0.5$. These are 
calculated for a stable top quark, that is ignoring the $b$-quark in the decay as a potential jet 
contributor. The absolute distribution (not shown here) therefore peaks around $\eta=\pm2.5$, as 
its dominant contribution is from the forward jet on the light line.

 Decreasing the
slicing cutoff $\taucutl$ exponentially by factors of 10, we see that the 
difference between $\taucutl = 0.001$ and $\taucutl = 0.01$ is negligible. Even for 
$\taucutl=0.1$ the difference is small, especially in 
the central region. This demonstrates the successful cancellation of \IR{} divergences through 
application of the slicing subtractions.
Note that the corrections in the center are negative, while they are positive in the tails. 
In the absence of any cuts this leads to strong cancellations and a small \NNLO{} correction 
to the inclusive cross section.

\subsection{Top-quark-decay corrections}

For the top-quark decay we make use of the factorization theorem described in 
refs.~\cite{Korchemsky:1994jb,Akhoury:1995fp,Bauer:2003pi,Bosch:2004th,Liu:2010ng} for the observable
$$\tau_d = p_X^2 / m_t^2\,,$$ 
where $p_X$ is the sum of all final-state parton momenta in the decay. The respective factorization 
theorem for corrections in decay and integrated over $\tau_d$ reads
\begin{multline}
\mathrm{d}\sigma_{ab}(\tau_d < \taucutd) = \mathrm{d}\sigma^\text{Born} f_a(x_a,\mu) f_b(x_b,\mu) \times
\\ 
H(x,\mu) \int_0^{\taucutd} 
\!\!\!\mathrm{d}\tau_d 
	\int\!\mathrm{d}m^2 \int\!\mathrm{d}k_s\, J(m^2,\mu)\, S(k_s,\mu) \, \delta\!\left( \tau_d - 
	\frac{m^2+2E_b k_s}{m_t^2} 
	\right)\,,
\end{multline}
where $x = Q^2 / m_t^2 > 0$, $E_b = (m_t^2-Q^2)/(2m_t)$ is the $b$-quark energy, and $Q^2 = 
m_W^2$ for an on-shell 
$W$-boson. We also allow for the generation of the $W$-boson decay distributed according to a
Breit-Wigner peak with $Q^2 \neq m_W^2$.

We use the bare soft function $S_\text{bare}$ from ref.~\cite{Becher:2005pd} and the bare jet 
function $J_\text{bare}$ from 
ref.~\cite{Becher:2006qw}. We combined these with the bare hard function $H_\text{bare}$, which has 
been independently calculated in four different references 
\cite{Bonciani:2008wf,Asatrian:2008uk,Beneke:2008ei,Bell:2008ws}\footnote{The hard function is 
meanwhile also available 
	with full bottom-quark mass dependence through ref.~\cite{Engel:2018fsb}.}. We perform all 
	integral 
convolutions in Laplace-space and find successful cancellation of all \IR{} poles between 
$J_\text{bare}$, $S_\text{bare}$ and $H_\text{bare}$. 
We furthermore checked that we can individually reproduce the renormalized results of the soft 
($S$) and jet functions ($J$) as given in refs.~\cite{Becher:2005pd,Becher:2006qw}.

For the hard function we successfully checked the agreement between all four calculations using the 
Mathematica package {\abbrev HPL} \cite{Maitre:2005uu} for the evaluation of harmonic 
polylogarithms.
We converted the hard function into an expression using only harmonic polylogarithms, so that our 
hard function 
implementation can be reused for the production part based on the automatic analytic continuation 
in the hplog Fortran 
code \cite{Gehrmann:2001pz}.

The matrix elements entering the above-cut calculation have been obtained
by crossing from the virtual amplitudes used in the computation of $Wt$
production at NLO~\cite{Campbell:2005bb} and the real-radiation amplitudes
employed in the four-flavor calculation of $t$-channel single-top
production~\cite{Campbell:2009ss}.  As for the corrections on the light-line,
we cross-checked all amplitudes by a numerical comparison with Recola \cite{Actis:2016mpe}.
We then set up the \NLO{} above-cut calculation using dipole subtractions.
For the case of $q \to q g$ splitting with the massive top-quark 
spectator we have used the dipole expression from ref.~\cite{Campbell:2004ch}, also presented in 
ref.~\cite{Melnikov:2011qx}\footnote{We note that eq.~5 in ref.~\cite{Melnikov:2011qx} contains a 
typo, where it should read $\eta\epsilon(1-z)$ instead of $y\epsilon(1-z)$, where $\eta$ is a 
parameter that distinguishes between the CDR and FDH schemes. }. The dipole expression for the $g \to g g$
splitting with the massive top spectator is taken from ref.~\cite{Melnikov:2011qx}.


\paragraph{Calculation of the top-quark decay width.}

\begin{figure}
	\centering
	\includegraphics{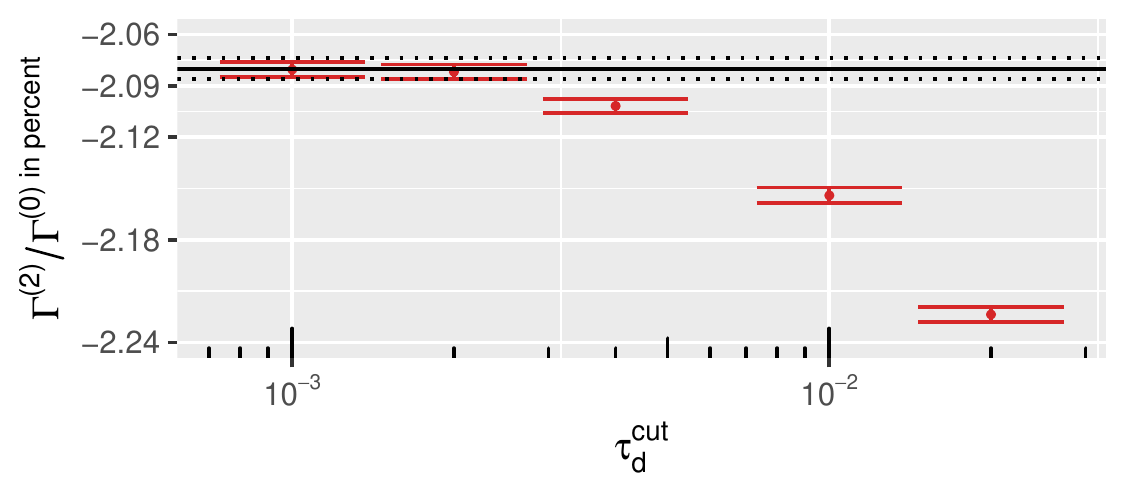}
	\caption{$\taucutd$-dependence of the top-quark decay width \NNLO{} corrections. Error-bars 
		indicate numerical integration uncertainties. The solid line (with dashed band) indicates the 
		fitted result (and uncertainties), which agrees with the result using $\taucutd=10^{-3}$. }
	\label{fig:taucut_width}
\end{figure}

Our calculation can be used to compute the \NNLO{} \QCD{} corrections to the
top-quark width.  Such a calculation has previously been performed
numerically in refs.~\cite{Gao:2012ja,Brucherseifer:2013iv} (\GLZ{}), as well as
analytically as an expansion in $m_W^2/m_t^2$~\cite{Blokland:2004ye,Blokland:2005vq} (\BCST{}).
We write the perturbative corrections to the top-quark width as $\Gamma_t = \Gamma_t^{(0)}(1 + 
\delta^{(1)} + \delta^{(2)})$, where 
the relative corrections of order $\alpha_s^k$ are written as 
$\delta^{(k)}\equiv\Gamma_t^{(k)}/\Gamma_t^{(0)}$ and $\Gamma_t^{(0)}$ 
is the \LO{} decay width.
In order to demonstrate the correctness of our calculation, in 
\cref{fig:taucut_width} we show the $\taucutd$-dependence of the \NNLO{} corrections 
$\delta^{(2)}$ for the choice $\mu=m_t$
\cref{fig:taucut_width}.  We observe a good convergence of the slicing procedure to the 
asymptotic value, finding no significant dependence around $\taucutd=10^{-3}$.

A comparison of our calculation with previous results for the top-quark width is shown in \cref{table:Gamma2},
demonstrating excellent agreement.  The previous approximate results are confirmed at
the 0.5\% level for the second-order coefficient, corresponding to a difference
of at most $0.01$\% in the total width.
The comparison is performed with $m_t=\SI{172.5}{\GeV}$, but the relative size of the
corrections is somewhat insensitive to the precise top-quark mass \cite{Gao:2012ja}. 

\begin{table}
	\centering
	\caption{Comparison between our calculation of the second-order correction to the
	top-quark width~($\delta^{(2)}$), as a percentage of the \LO{} result of
	$\SI{1.4806}{\GeV}$ ($m_t=\SI{172.5}{\GeV}$), and
	previous computations of the same quantity from Gao et 
	al.~\cite{Gao:2012ja}~($\delta^{(2)}_\GLZ{}$)
	and Blokland et al.~\cite{Blokland:2005vq}~($\delta^{(2)}_\BCST{}$).}
	\vspace*{0.5em}
	\bgroup
		\setlength\tabcolsep{0.5em}
	\def\arraystretch{1.5}%
	\begin{tabular}{@{}c|ccc@{}}
		\toprule
		$\mu$ &   $\delta^{(2)}_\text{BCST}$ &  $\delta^{(2)}_\text{GLZ}$ &  $\delta^{(2)}$ \\
 \midrule
		$m_t$     & $-2.07$ & $-2.08$ & $-2.08(1)$ \\

		$2 m_t$   & $-2.38$ & $-2.39$ & $-2.39(1)$ \\

		$m_t / 2$ & $-1.58$ & $-1.58$ & $-1.58(1)$ \\
\bottomrule
	\end{tabular}
	\egroup
	\label{table:Gamma2}
\end{table}

\subsection{Top-quark-production corrections on the heavy line}

For the heavy line production corrections we make use of the factorization theorem given in 
ref.~\cite{Berger:2016inr}
using the observable $\tau_h$ \cite{Stewart:2010tn},
$$ \tau_h = \frac{2 p_X p_b}{m_t^2 - q^2}\,,$$
where $p_X$ is the sum of the additionally radiated partons, $p_b$ is the $b$-quark beam 
momentum and $q^2$ is the momentum-squared of the $t-$channel $W$ boson ($q^2 < 0$).

The factorization theorem in $\tau_h$, integrated over $\tau_h$, reads
\begin{multline}
\mathrm{d}\sigma_{ab}(\tau_h < \taucuth) = \mathrm{d}\sigma^\text{Born} f_a(x_a,\mu)\, \times \\
H(x+i0,\mu)  \int_0^{\tau_h^\text{cut}}\!\!\!\mathrm{d}\tau_h 
\int \!
\mathrm{d}t \int\!\mathrm{d}k_s\, 
B_b(x_b,t,\mu)\, S(k_s,\mu,\mu_F)\, \delta\!\left( \tau_h - 
\frac{t+2 k_s E_b}{m_t^2 - 
q^2}  
\right)\,,
\end{multline}
where $x = q^2 / m_t^2$.  $E_b=(m_t^2-q^2)/(2m_t)$ is the $b$-quark beam energy in 
the top-quark rest 
frame. We reuse the renormalized soft function $S$ from the top-quark decay 
\cite{Becher:2005pd}. This is possible when the soft function is defined in the top-quark rest 
frame as above and has been discussed in ref.~\cite{Berger:2017zof}. We also reuse the
beam functions $B_b$ \cite{Stewart:2009yx,Stewart:2010tn,Stewart:2010qs} implemented through 
ref.~\cite{Boughezal:2016wmq}. 
The hard function $H$ is reused from the top-quark decay, but requires an analytic continuation. 
Since we implemented it in terms of harmonic polylogarithms, we can use the automatic analytic 
continuation of the hplog Fortran code \cite{Gehrmann:2001pz}. We have verified that 
these analytically continued numerical 
results agree with an explicit analytic continuation in Mathematica using the {\abbrev HPL} package 
\cite{Maitre:2005uu}.  Lastly, we have explicitly performed the integrals and distributional 
convolutions 
in Laplace space to obtain the assembled result in terms of $\log^k(\tau_h)$ contributions.
As for the other contributions, we analytically verified that the renormalization scale dependence 
vanishes to the corresponding order, or in other words, that all 
poles cancel between bare hard, soft and beam function. We also find full agreement 
between our $\tau_h$-subtraction results at \NLO{} and a calculation using dipole subtractions 
\cite{Campbell:2004ch}. The above-cut contributions are obtained by crossing from the matrix 
elements used in the decay corrections.

\begin{figure}
	\centering
	\includegraphics{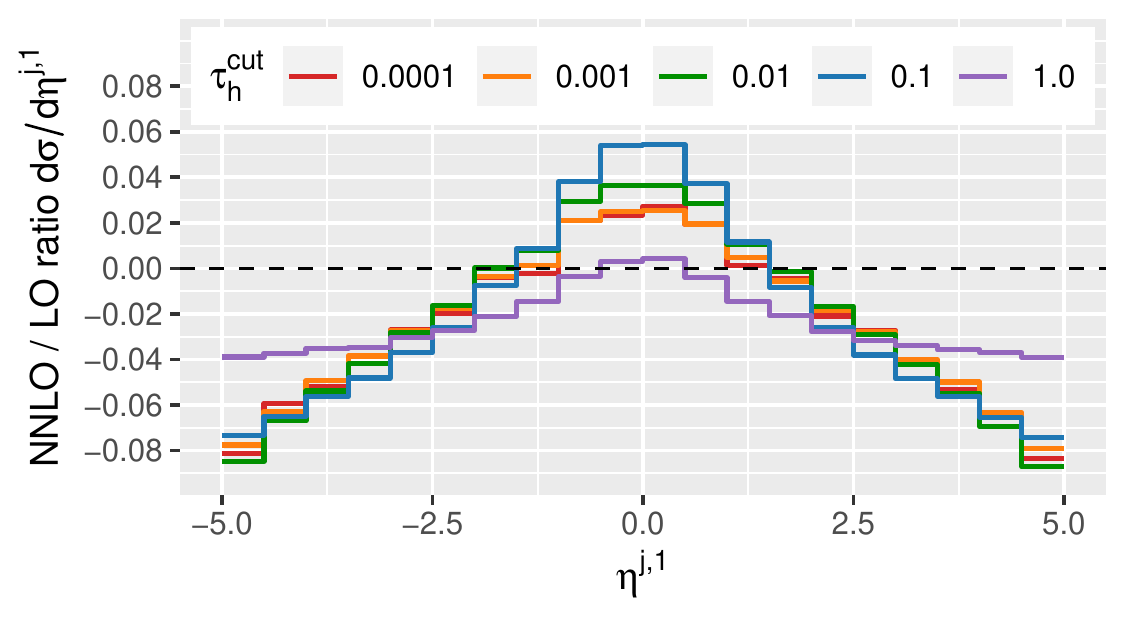}
	\caption{\NNLO{} corrections from the heavy-line production vertex for the leading jet 
	pseudorapidity distribution, relative to 
	\LO{}. The (automatically) fitted result coincides within numerical uncertainties with the 
	$\taucuth=10^{-3}$ result.}
	\label{fig:heavyetaj1}
\end{figure}

We show the \NNLO{} corrections
to the leading jet pseudorapidity distribution as a fraction of the \LO{} result in 
\cref{fig:heavyetaj1}, as well as their dependence on $\taucuth$. As for the light-line corrections, this 
distribution is calculated for a stable top quark, thereby ignoring the $b$-quark in the decay as a 
jet contributor. Comparing the results for the exponentially decreasing slicing cutoffs $\taucuth$, 
we see that $\taucuth=10^{-4}$ and $\taucuth=10^{-3}$ differ only marginally within remaining 
numerical uncertainties. Even $\taucuth=10^{-2}$ differs only in the most central bins.
This demonstrates the successful cancellation of \IR{} divergences.
We conclude that $\taucuth=0.001$ constitutes a sufficiently small cutoff for reliable results.
Once more there are strong cancellations, that lead to a small correction to the inclusive cross 
section, since the corrections are small and positive in the central region and larger and negative 
towards the tails.

\subsection{$\NLO{}\otimes\NLO{}$ interference contributions }

Apart from two-loop vertex corrections, $\NLO{}\otimes\NLO{}$ contributions also arise at \NNLO{}
from production corrections times decay corrections (light\,$\otimes$\,decay, 
heavy\,$\otimes$\,decay) 
and light-line corrections times heavy-line corrections (light\,$\otimes$\,heavy).

We implemented all of these contributions in a mixed scheme where one part of the \NLO{} 
calculation is handled with dipole subtractions and the other half with \NLO{} slicing 
subtractions, as described for the individual \NNLO{} vertex corrections above. This 
makes the 
implementation within our existing 
infrastructure easier. Each of the three $\NLO{}\otimes\NLO{}$ calculations 
(light\,$\otimes$\,decay, 
heavy\,$\otimes$\,decay, light\,$\otimes$\,heavy) consists of four contributions,
categorized into real\,$\otimes$\,real, (RR), real\,$\otimes$\,virtual (RV), 
virtual\,$\otimes$\,real (VR), 
and
virtual\,$\otimes$\,virtual (VV). The first part is always calculated with dipole subtractions and 
denotes the 
real emission contributions with dipole subtractions (real), and virtual loop corrections with 
integrated dipoles (virtual). The second part is calculated using the discussed factorization 
theorems 
and denotes the above-cut emission (real), and the below-cut loop contribution (virtual) from the 
integrated factorization theorem.

For each of the three $\NLO{}\otimes\NLO{}$ calculations we checked cancellation of poles
for the dipole subtractions through the alpha parameter 
\cite{Nagy:1998bb,Nagy:2003tz,Campbell:2004ch,Campbell:2005bb,Bevilacqua:2009zn} for the RV+VV and 
RR+VR contributions 
separately. We furthermore checked for VV+VR and RV+RR separately that remaining $\taucut$ effects 
for our default choice of $\taucut=10^{-3}$ are negligible at the per-mille level and that the 
asymptotic approach is 
logarithmic as predicted by the factorization theorems.
Therefore all four pieces of each $\NLO{}\otimes\NLO{}$ 
contribution are checked through a chain of tests.

\begin{figure}
	\centering
	\includegraphics{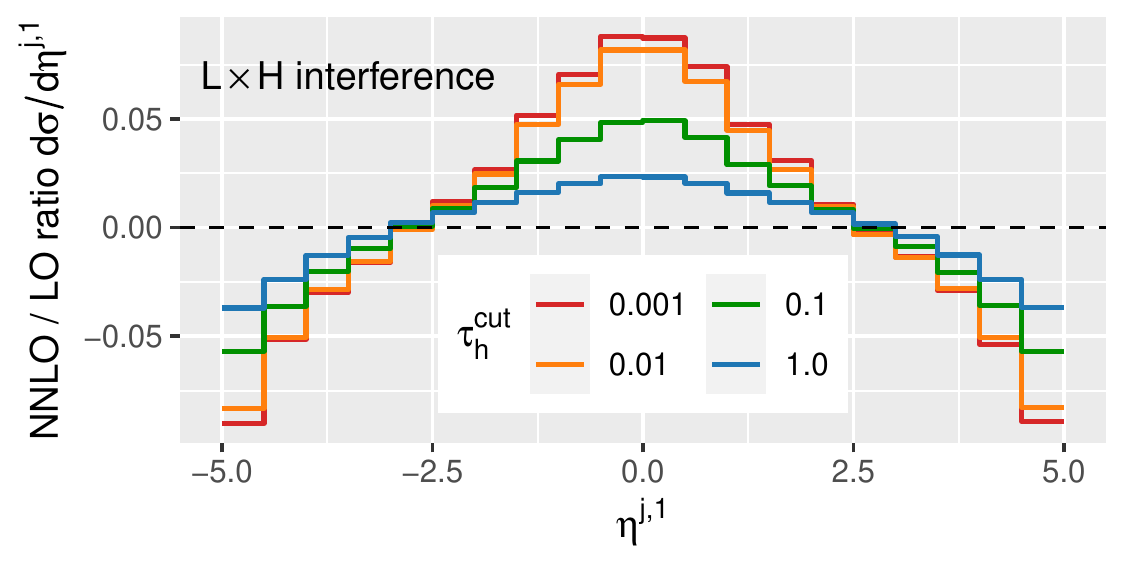}
	\includegraphics{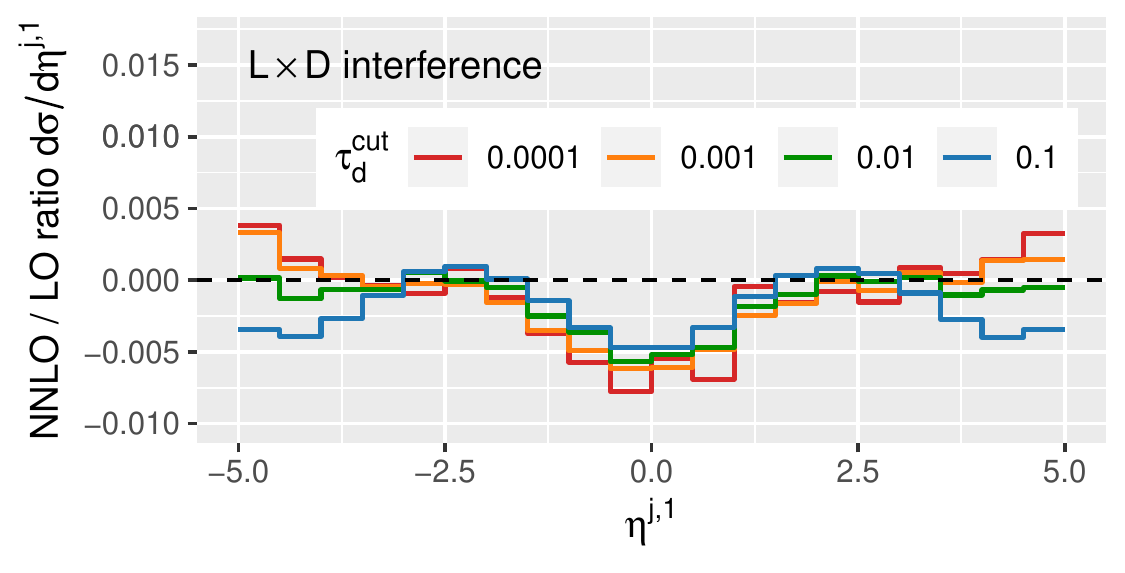}
	\includegraphics{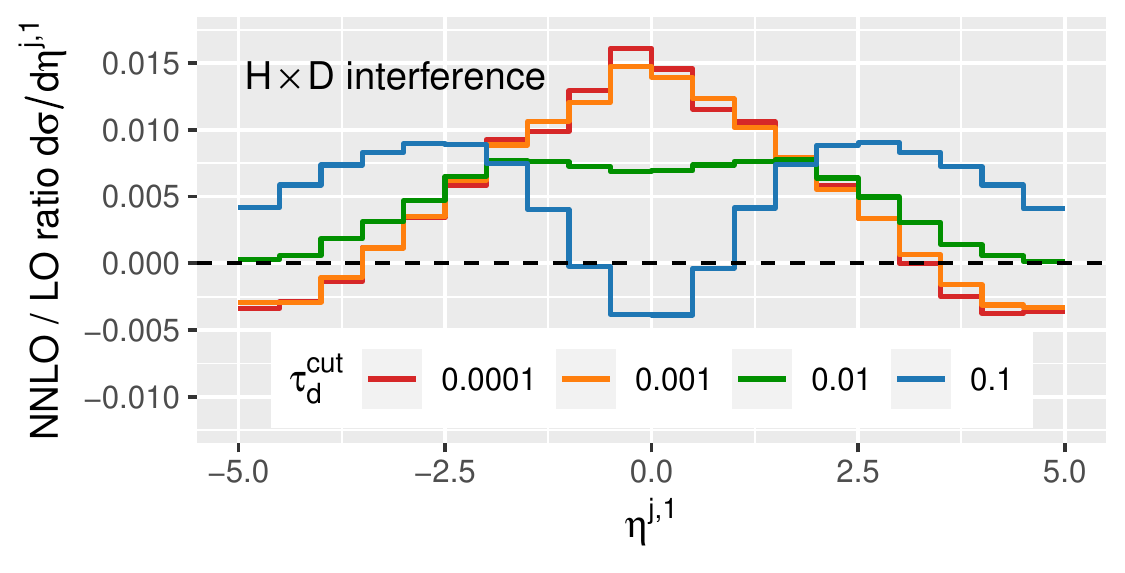}
	\caption{\NNLO{} corrections to the leading 
	jet pseudorapidity distribution, relative to \LO{}, from light\,$\otimes$\,heavy interference (upper panel),
	light\,$\otimes$\,decay interference (center panel) and heavy\,$\otimes$\,decay interference (lower panel).}
	\label{fig:taucut_interference}
\end{figure}

We show the differential $\taucut$-dependence 
for all three interference contributions in \cref{fig:taucut_interference} for 
the leading jet pseudorapidity distribution at $\sqrt{s}=\SI{13}{\TeV}$ using anti-$k_T$ jet 
clustering with $R=0.5$. We indeed find that for 
$\taucut=10^{-3}$ residual power corrections are negligible at 
the single per-mille level.

While one might
naively expect that the factorization between production and decay allows an implementation of \NNLO{}
interference corrections as the 
product of \NLO{} corrections in production and decay, at the level of differential cross sections, 
this is not true. This property only 
holds fully inclusively, and serves as another check of the
implementation, see \cref{eq:fixedorderexpansion}. 
For instance, the light\,$\otimes$\,decay interference 
contribution ($\sigma_{L \times D}$) is equal to the light-line 
production \NLO{} coefficient ($\sigma^{(1)}_{L}$) multiplied  by
the \NLO{} width correction factor, $\Gamma^{(1)}/\Gamma^{(0)}=-0.0858$.
As an example, at $\sqrt{s}=\SI{13}{\TeV}$ we have,
\begin{equation}
\sigma_{L \times D} = \SI{-154(4)}{fb}\,, \qquad
\sigma^{(1)}_{L} \times \frac{\Gamma^{(1)}}{\Gamma^{(0)}} = \SI{-155}{fb} \,.
\end{equation}
Furthermore, the heavy\,$\otimes$\,decay interference ($\sigma_{H \times D}$)
is also in excellent agreement,
\begin{equation}
\sigma_{H \times D} = \SI{645(4)}{fb}\,, \qquad
\sigma^{(1)}_{H} \times \frac{\Gamma^{(1)}}{\Gamma^{(0)}} = \SI{645}{fb} \,.
\end{equation}

\section{Comparison with previous results}
\label{sec:comparison}

In \cref{sec:calculation} we have presented extensive cross-checks of all parts of our
calculation, ranging from the amplitudes, to analytical checks of the factorization theorems and 
their ingredients, to the 
numerical implementation. Thus, having confidently assessed the validity of our calculation, we now 
compare against the two previous calculations of this process.  Like our calculation,
these are also performed in the approximation of an on-shell top quark,
but the previous results disagree on the size of the \NNLO{} corrections.

\subsection{Stable top quark}
\label{sec:comparisonBCM}

We first compare with the \NNLO{} results computed by Brucherseifer, Caola and Melnikov
(\BCM{}) \cite{Brucherseifer:2014ama}. This calculation is performed in the stable-top-quark 
approximation and therefore does not include the decay of the top 
quark. To compare with their results we adopt
$\sqrt{s}=\SI{8}{\TeV}$, $m_t=\SI{173.2}{\GeV}$, $m_W=\SI{80.398}{\GeV}$, 
$G_F=\SI{1.16639d-5}{\GeV^{-2}}$ and use the {\abbrev MSTW2008} \PDF{} set \cite{MSTW2008} that
corresponds to the order of the calculation. 
The resulting comparison between our calculation and the
results of ref.~\cite{Brucherseifer:2014ama} is shown in \cref{table:BCM-top},
where we tabulate cross sections for four different minimum top-quark transverse momenta ($p_T$),
including uncertainties due to scale variation.

While we find agreement at the per-mille level throughout \NLO{}, we find discrepancies at the 
level of about $1$ to $1.5$ percent for the \NNLO{} results. The calculational 
uncertainties 
for both 
our 
and 
the 
\BCM{} results are about $\SI{0.1}{pb}$, which is a relative uncertainty of two 
per-mille.\footnote{We assume that for \BCM{} results the uncertainties are one in the last 
significant digit of the quoted 
result.}
A first re-calculation of single-top-quark production in refs.~\cite{Berger:2016oht,Berger:2017zof}
by Berger, Gao, Zhu (\BGZ{}) has reported a similarly-sized difference to the \BCM{} calculation. 
While these authors did not perform a direct comparison with the \BCM{} results, we will
demonstrate in \cref{sec:comparisonBGZ} a comparison with the 
results of \BGZ{} that finds agreement at \NNLO{} at a much greater level (within a few 
per-mille), even for fiducial and differential results.

\begin{table}
	\centering
	\caption{Comparison with fully inclusive top-quark production results from
	Brucherseifer, Caola, Melnikov in ref.~\cite{Brucherseifer:2014ama}. Cross-sections in 
	picobarns. Scale uncertainties in super- and subscript are from simultaneous variation of 
	$\mu_R=\mu_F=m_t$ by a factor of two and one half, respectively.}
	\vspace*{0.5em}
	\bgroup
		\setlength\tabcolsep{0.5em}
	\def\arraystretch{1.5}%
	\begin{tabular}{@{}l|cc|cc|cc@{}}

		\toprule
		$p_{T,\text{min}}^\text{top}$                                & 
		$\sigma_\text{LO}^\text{BCM}$ & 
		$\sigma_\text{LO} \pm 0.01$ & $\sigma_\text{NLO}^\text{BCM}$ & $\sigma_\text{NLO} \pm 0.05$ 
		& 
		$\sigma_\text{NNLO}^\text{BCM}$ & $\sigma_\text{NNLO} \pm 0.1$ \\
 \midrule
		\SI{0}{\GeV}  & $53.8^{+3.0}_{-4.3}$ & $53.79^{+3.03}_{-4.33}$ & $55.1^{+1.6}_{-0.9}$  & 
		$55.15^{+1.63}_{-0.90}$ & \boldmath$54.2^{+0.5}_{-0.2}$ & $\mathbf{53.5^{+0.65}_{-0.31}}$ \\

		\SI{20}{\GeV} & $46.6^{+2.5}_{-3.7}$ & $46.65^{+2.53}_{-3.66}$ & $48.9^{+1.2}_{-0.5}$  & 
		$48.90^{+1.22}_{-0.54}$ & \boldmath$48.3^{+0.3}_{-0.02}$ & $\mathbf{47.9^{+0.44}_{-0.16}}$ 
		\\

		\SI{40}{\GeV} & $33.4^{+1.7}_{-2.5}$ & $33.41^{+1.67}_{-2.48}$ & $36.5^{+0.6}_{-0.03}$  & 
		$36.55^{+0.59}_{-0.03}$ & $36.5^{+0.1}_{+0.1}$ & ${36.5^\mathbf{+0.15}_\mathbf{+0.04}}$ \\

		\SI{60}{\GeV} & $22.0^{+1.0}_{-1.5}$ & $21.95^{+0.99}_{-1.52}$ & $25.0^{+0.2}_{+0.3}$  & 
		$25.08^{+0.17}_{-0.27}$ & $25.4^{-0.1}_{+0.2}$ & $25.4^{\mathbf{+0.03}}_{\mathbf{+0.13}}$ 
		\\        
\bottomrule
	\end{tabular}
	\egroup
	\label{table:BCM-top}
\end{table}

\begin{figure}
	\centering
	\includegraphics{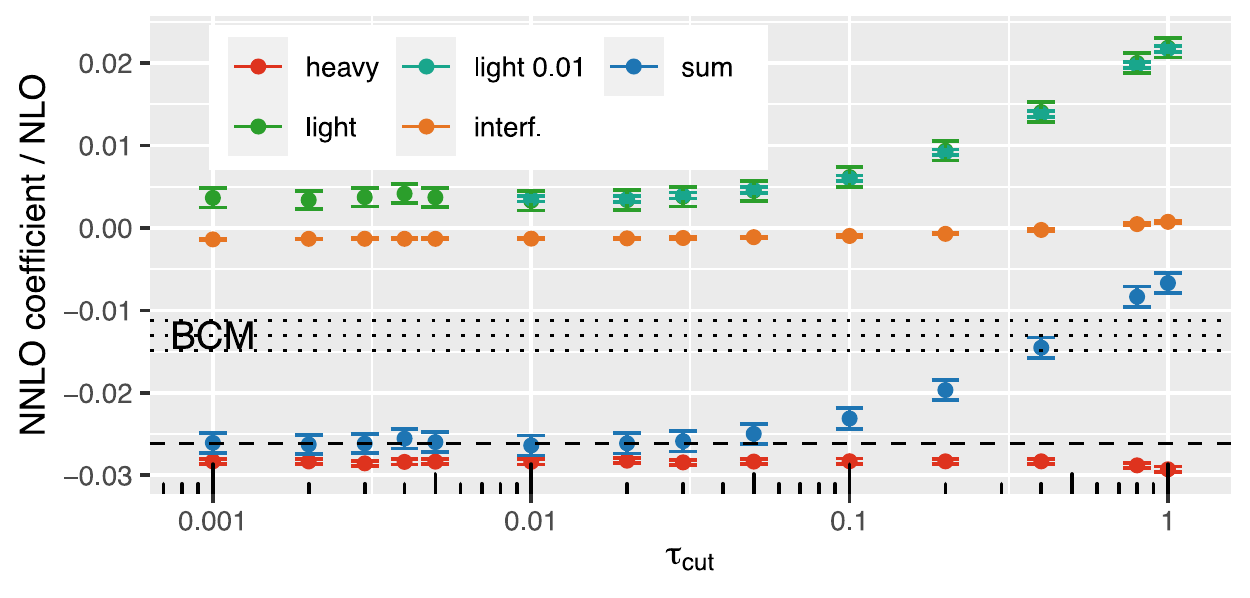}
	\caption{$\taucut$ dependence of \NNLO{} light, heavy, interference and summed contributions 
	relative to the \NLO{} part for the fully inclusive result in \cref{table:BCM-top}. The dotted 
	area represents the \NNLO{} coefficient with an absolute uncertainty of $\pm\SI{0.1}{pb}$ to 
	obtain the result from Brucherseifer, Caola, Melnikov (BCM) \cite{Brucherseifer:2014ama}.}
	\label{fig:taucut_melnikov}
\end{figure}

The numerical integration uncertainties and residual systematic uncertainty from our slicing
cutoff $\taucut$ are at the per-mille level and small compared to the discrepancy we find. We first
demonstrate this in the fully inclusive case ($p^\text{top}_{T,\text{min}}=0$) in 
\cref{fig:taucut_melnikov}.
We show the $\taucut$ dependence of the heavy-line contribution, light-line contribution,
interference contribution and the sum, relative to the \NLO{} part of the \NNLO{} result. 
The black dotted area represents the (reconstructed) \NNLO{} coefficient of \BCM{} shown in 
\cref{table:BCM-top} with an absolute uncertainty of $\pm \SI{0.1}{pb}$. Our numerical
integration uncertainty is indicated by the error bars. All results have been obtained
using the multi-$\taucut$ sampling in \MCFM{}-9 with a minimum $\taucut$ of $10^{-3}$. Since
the largest uncertainties are for the light-line contributions, the figure also
shows a separate re-calculation of the light-line corrections with a nominal $\taucutl$ value of
$0.01$. These have a significantly smaller integration uncertainty 
but agree perfectly with results calculated with a nominal value of $\taucut=0.001$.
Overall, with $\taucut=0.01$ we are already well within the asymptotic regime where results are 
precise at the per-mille level. The discrepancy with the \BCM{} result is clearly visible and
cannot be explained by calculational uncertainties or a difference in the \NLO{} contribution.

\begin{figure}
	\centering
	\includegraphics{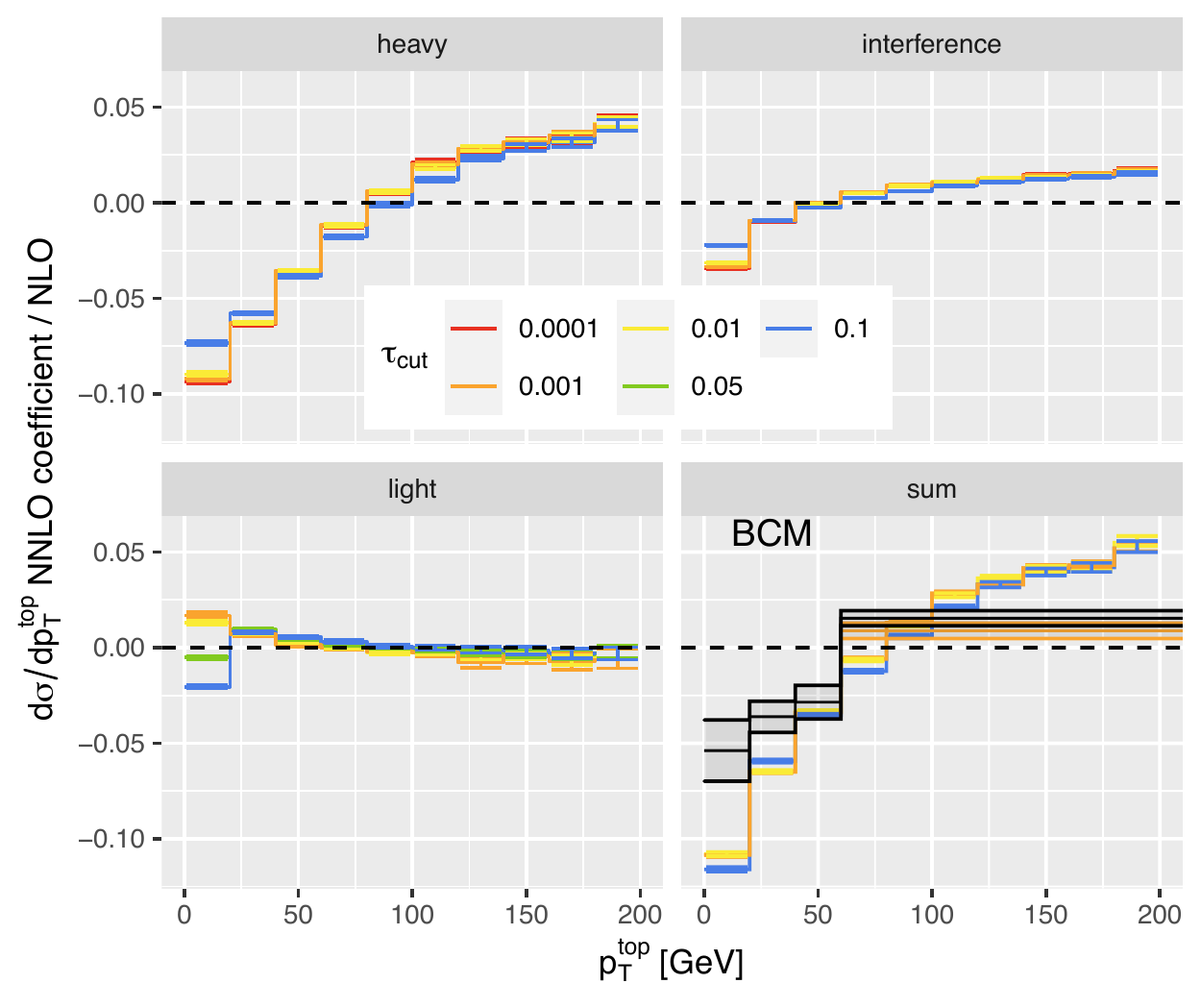}
	\caption{$\taucut$ dependence of \NNLO{} light, heavy, interference and summed contributions 
	relative to the \NLO{} part for the top-quark $p_T$ distribution as input to 
	\cref{table:BCM-top}. The black points with error bars represent the reconstructed results from 
	Brucherseifer, Caola, Melnikov (BCM) \cite{Brucherseifer:2014ama}, see \cref{table:BCM-top}, 
	with an absolute uncertainty of $\pm\SI{0.1}{pb}$.}
	\label{fig:taucut_pttop_melnikov}
\end{figure}

We next consider the top-quark $p_T$ distribution which, when integrated or summed over bins, 
reproduces the numbers in \cref{table:BCM-top} that require a minimum top-quark $p_T$. Also here
we have to carefully inspect the $\taucut$ dependence, since it may manifest in a different
way depending on the kinematics.
For this we present top-quark $p_T$-differential results in \cref{fig:taucut_pttop_melnikov}
for $\taucut$ values varying over three orders of magnitude. The results are presented normalized
to the \NLO{} part of the distribution, that is to the \NLO{} cross section calculated with \NNLO{} 
\PDF{}s, to emphasize the relative size of the \NNLO{} contributions. Just as in the fully 
inclusive case it is evident that with $\taucut=0.01$ the asymptotic regime is reached for all 
contributions, since it agrees well within
our numerical uncertainties (small error bars) with the $\taucut=0.001$ result. For comparison
with \BCM{} we have reconstructed their corresponding results for the first three bins from 
\cref{table:BCM-top} and assume an absolute uncertainty of $\pm \SI{0.1}{pb}$. 

Our results show a definitive difference with the \BCM{} results at both the differential and inclusive 
level. At the total level the discrepancy is about one percent. The discrepancy with the \NNLO{}
coefficient itself is 100\%: For fully inclusive top-quark production our \NNLO{} coefficient is
$-\SI{1.4}{pb}$, while the \BCM{} coefficient is just $-\SI{0.7}{pb}$. 
At the differential level the discrepancy is also large. For example for the top-quark $p_T$ 
distribution the difference in the first bin from  \SIrange{0}{20}{\GeV} is six percent relative to 
\NLO{}. For the last reconstructed bin
for $p_T>\SI{60}{\GeV}$ we agree within mutual uncertainties. We
observe similarly-sized discrepancies for an anti-top quark, as detailed in  \cref{table:BCM-anti-top}. 

\begin{table}
	\centering
	\caption{Comparison with fully inclusive anti-top-quark production results from
		Brucherseifer, Caola, Melnikov in ref.~\cite{Brucherseifer:2014ama}. Cross-sections in 
		picobarns. Scale uncertainties in super- and subscript from simultaneous variation of 
		$\mu_R=m_t$ and $\mu_F=m_t$ by a factor of two and one half, respectively.}
	\vspace*{0.5em}
	\bgroup
	\setlength\tabcolsep{0.5em}
	\def\arraystretch{1.5}%
	\begin{tabular}{@{}l|cc|cc|cc@{}}
		
		\toprule
		$p_{T,\text{min}}^\text{anti-top}$                                & 
		$\sigma_\text{LO}^\text{BCM}$ & 
		$\sigma_\text{LO} \pm 0.01$ & $\sigma_\text{NLO}^\text{BCM}$ & $\sigma_\text{NLO} \pm 0.05$ 
		& 
		$\sigma_\text{NNLO}^\text{BCM}$ & $\sigma_\text{NNLO} \pm 0.1$ \\
		\midrule
		\SI{0}{\GeV}  & $29.1^{+1.7}_{-2.4}$ & $29.06^{+1.67}_{-2.38}$ & $30.1^{+0.9}_{-0.5}$  & 
		$30.12^{+0.91}_{-0.48}$ & \boldmath$29.7^{+0.3}_{-0.1}$ & \boldmath$29.2^{+0.37}_{-0.19}$ \\
		
		\SI{20}{\GeV} & $24.8^{+1.4}_{-2.0}$ & $24.78^{+1.36}_{-1.97}$ & $26.3^{+0.7}_{-0.3}$  &
		$26.39^{+0.65}_{-0.26}$ & \boldmath$26.2^{-0.01}_{-0.1}$ & \boldmath$25.9^{+0.24}_{-0.10}$ 
		\\
		
		\SI{40}{\GeV} & $17.1^{+0.9}_{-1.3}$ & $17.12^{+0.86}_{-1.28}$ & $19.1^{+0.3}_{+0.1}$  &
		$19.20^{+0.27}_{+0.05}$ & $19.3^{-0.2}_{+0.1}$ & 	
		$19.2^{\mathbf{+0.07}}_{\mathbf{+0.02}}$ \\
		
		\SI{60}{\GeV} & $10.8^{+0.5}_{-0.7}$ & $10.78^{+0.48}_{-0.75}$ & $12.7^{+0.03}_{+0.2}$  &
		$12.73^{+0.03}_{+0.20}$ & $12.9^{-0.2}_{+0.2}$ & $13.0^{\mathbf{-0.03}}_{\mathbf{+0.07}}$ 
		\\        
		\bottomrule
	\end{tabular}
	\egroup
	\label{table:BCM-anti-top}
\end{table}

\subsection{Top quark production and decay}
\label{sec:comparisonBGZ}

We now turn to a comparison with the re-calculation by Berger, Gao, Zhu 
(\BGZ{}) \cite{Berger:2016oht,Berger:2017zof}, who also include the top-quark decay and provide 
fully fiducial predictions and differential results with which to compare.

We first compare with inclusive cross sections (table 1 in ref.~\cite{Berger:2017zof}), at $\sqrt 
s=7$~TeV and $\sqrt s=14$~TeV, for $t$ and $\bar t$. This comparison is performed with 
$m_t=\SI{172.5}{\GeV}$, $m_W=\SI{80.385}{\GeV}$, $G_F=\SI{1.166379d-5}{\GeV^{-2}}$ and the {\abbrev 
CT14nnlo} \PDF{} set \cite{CT14}, irrespective of the order of the calculation that is considered. At \NNLO{} we 
obtain all 
results with a nominal $\taucut=0.001$ but ensure through the multi-$\taucut$ sampling in 
\MCFM{}-9 that the residual $\taucut$ dependence is small compared to our quoted combined 
integration and residual $\taucut$ 
uncertainty. The perturbative truncation uncertainties from scale variation are obtained by varying 
$\mu_R=\mu_F=m_t$ simultaneously by a factor of two and one half, respectively, so by a two-point 
variation. We do this to compare with the results in ref.~\cite{Berger:2017zof}, but note that with 
a six-point variation, where $\mu_R$ and $\mu_F$ are varied independently, the uncertainties 
increase at both \NLO{} and \NNLO{}.
The results in \cref{table:BGZ-inclusive} show that, for both the central values and scale
variations, the two calculations are in agreement at the level of less than three per-mille in the 
\NNLO{} cross sections. Since our overall numerical and calculational precision is 
at the level of two to three per-mille, we consider this to be a excellent agreement, which is 
further supported by the following fiducial comparison.

\begin{table}
	\centering
	\caption{Comparison with fully inclusive production results from
		Berger, Gao, Zhu \cite{Berger:2017zof,Berger:2016oht}. Scale uncertainties in super- and 
		subscript from simultaneous variation of $\mu_R=m_t$ and $\mu_F=m_t$ by a factor of two and 
		one half, respectively.}
	\vspace*{0.5em}
	\bgroup
	\setlength\tabcolsep{0.5em}
	\def\arraystretch{1.5}%
	\begin{tabular}{@{}ll|ll|ll|ll@{}}
		
		\toprule

		&          &  $\sigma_\text{LO}^\text{BGZ}$  & $\sigma_\text{LO}$ & 
		$\sigma_\text{NLO}^\text{BGZ}$ & $\sigma_\text{NLO} \pm 0.01$ & 
		$\sigma_\text{NNLO}^\text{BGZ}$ & $\sigma_\text{NNLO}$ \\ \midrule

		\multirow{2}{*}{\SI{7}{\TeV}} \hspace{-0.5em} & top      & $44.55^{+5.3\%}_{-7.5\%}$   & 
		$44.55^{+5.3\%}_{-7.5\%}$   & $43.14^{+2.9\%}_{-1.6\%}$    &  $43.15^{+2.9\%}_{-1.6\%}$   
		&  $42.05^{+1.2\%}_{-0.6\%}$    & $41.99(4)^{+1.4\%}_{-0.7\%}$    \\

		& anti-top & $23.29^{+5.3\%}_{-7.6\%}$   &  $23.29^{+5.3\%}_{-7.6\%}$  & 
		$22.57^{+2.9\%}_{-1.5\%}$     & $22.57^{+2.9\%}_{-1.5\%}$    & 
		$21.95^{+1.2\%}_{-0.7\%}$     & $21.90(3)^{+1.4\%}_{-0.8\%}$    \\
\midrule
		\multirow{2}{*}{\SI{14}{\TeV}} \hspace{-0.5em} & top      & $164.4^{+8.4\%}_{-10\%}$   &  
		$164.41^{+8.4\%}_{-10.6\%}$    & $157.8^{+3.0\%}_{-1.7\%}$    & 
		$157.78^{+2.9\%}_{-1.7\%}$    & $153.3^{+1.1\%}_{-0.5\%}$     & 
		$153.2(2)^{+1.2\%}_{-0.6\%}$     \\

		& anti-top &  $99.60^{+8.7\%}_{-11\%}$  &   $99.60^{+8.7\%}_{-10.9\%}$  & 
		$94.77^{+3.0\%}_{-1.6\%}$    &  $94.77^{+3.0\%}_{-1.7\%}$   &  $91.81^{+1.0\%}_{-0.5\%}$    
		& $\enspace91.5(1)^{+1.2\%}_{-0.7\%}$    \\
				\bottomrule
	\end{tabular}
	\egroup
	\label{table:BGZ-inclusive}
\end{table}

\begin{figure}
	\centering
	\includegraphics{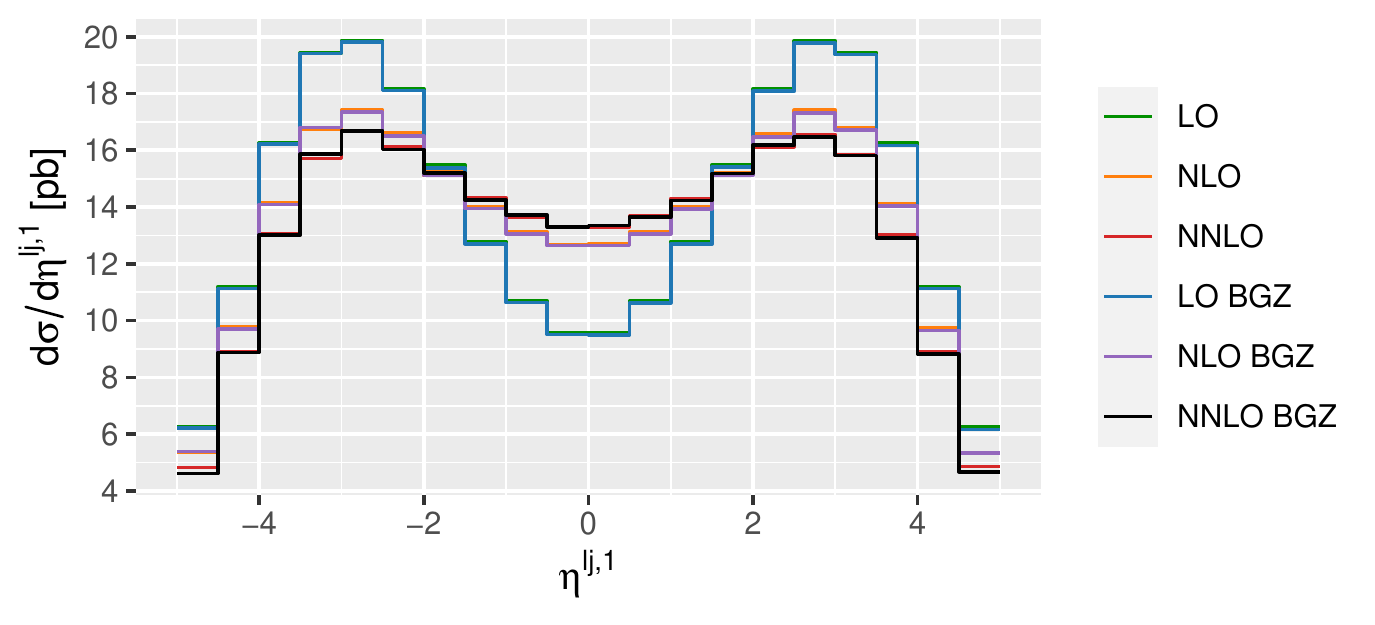}
	\caption{Leading jet pseudorapidity distribution in stable top-quark production at \LO{}, 
	\NLO{} and \NNLO{} using 
	\NNLO{} \CTFOURTEEN{} \PDF{}s. Our cross sections are compared with the results 
	in 	ref.~\cite{Berger:2017zof} (\BGZ{}).}
	\label{fig:eta_lj1_sum_bgz}
\end{figure}

We also successfully compared at the differential level for the fully inclusive process (stable top 
quark). For example, we show the leading jet pseudorapidity distribution in 
\cref{fig:eta_lj1_sum_bgz} where agreement within numerical uncertainties can be seen.
Although not illustrated explicitly here, the same holds true for the individual components of the
calculation:  heavy-line corrections, light-line corrections and light\,$\otimes$\,heavy interference
contributions.\footnote{We thank Jun Gao for providing the corresponding division of the \NNLO{} result.}

\paragraph{Comparison with fiducial results.}

\begin{table}
	\centering
	\caption{Kinematical cuts at $\sqrt{s}=\SI{13}{\TeV}$ for comparison with \BGZ{}, 
		ref.~\cite{Berger:2017zof}. The top-quark mass is $\SI{173.3}{\GeV}$. }
	\vspace*{0.5em}
	\bgroup
	\setlength\tabcolsep{1em}
	\def\arraystretch{1.5}%
	\begin{tabular}{l|c}
		Lepton cuts & $p_T^{l} > \SI{30}{\GeV}, \abs{\eta^l} < 2.4$\\
		anti-$k_T$ jet clustering & $p_T^{j} > \SI{40}{\GeV}, R=0.5, \abs{\eta_j}<5$\\
		jet requirements &	exactly two jets; with least one $b$ jet and $p_T$-leading $b$ jet 
		$\abs{\eta}<2.4$.
	\end{tabular}
	\egroup
	\label{tab:fiducialcuts_BGZ}
\end{table}

We now compare with the fiducial results in ref.~\cite{Berger:2017zof} (\BGZ{}), where we adopt
$m_t=\SI{173.3}{\GeV}$ and the kinematical cuts are summarized in \cref{tab:fiducialcuts_BGZ}.

The results of this comparison are presented in \cref{table:BGZ-fiducial}.  Through \NNLO{} we find 
agreement for all contributions within mutual uncertainties, where we assume an uncertainty of one 
in the final digit of the \BGZ{} results. In addition to this we have performed a check
of the individual contributions to the top-quark production number of \SI{-0.24}{pb} in \cref{table:BGZ-fiducial} 
from corrections to the light-line, heavy-line and heavy-light interference.
For the 
light-line contribution we find \SI{-100(2)}{fb}, in agreement with \SI{-100}{fb} (\BGZ{}), and
the interference contribution is identical, \SI{+135(1)}{fb} compared to \SI{+135}{fb} (\BGZ{}).
However, the 
heavy-line production contribution differs slightly: we find \SI{-260(2)}{fb}, to be compared with \SI{-273}{fb} (\BGZ{}).
While this tension is relatively large on the \NNLO{} heavy-line production coefficient itself, it is only a per-mille 
level effect on the full \NNLO{} result and does not affect the overall level of agreement.
Subsequent communication regarding this comparison revealed a small error in the earlier published calculation that
has now been identified and this small discrepancy is now understood.\footnote{We thank the authors of
Ref.~\cite{Berger:2017zof} for providing the detailed breakdown discussed here, and for clarifying this point.}

\begin{table}
	\centering
	\caption{Comparison with fiducial results in ref.~\cite{Berger:2017zof,Berger:2016oht}. Cross 
	sections are given in pb.}
	\vspace*{0.5em}
	\bgroup
	\setlength\tabcolsep{0.5em}
	\def\arraystretch{1.5}%
\begin{tabular}{@{}ll|ll|ll|ll@{}}

	\toprule

	&           & $\sigma_\text{LO}^\text{BGZ}$ & $\sigma_\text{LO}$ & 
	$\sigma_\text{NLO}^\text{BGZ}$ & $\sigma_\text{NLO}$ & $\sigma_\text{NNLO}^\text{BGZ}$ & 
	$\sigma_\text{NNLO}$ \\ \midrule

	\multirow{7}{*}{top}      & \textit{total}     & $4.067$   &   
	$4.07$  &  $2.95$   &  
	$2.94$    & $2.70$     & 
	$2.70$     \\
\cmidrule(l){2-8} 
	& production     &    &    & $-0.79$     & $-0.792(1)$    & $-0.24$     & 
	$-0.225(3)$     \\
\cmidrule(l){2-8} 
	& decay     &    &    & $-0.33$ &  $-0.338(1)$   &  $-0.13$    & $-0.126(1)$     \\
\cmidrule(l){2-8} 
	& prod.~$\times$~decay &    &    &     &     & $+0.12$     &  $+0.117(1)$    \\
\midrule
	\multirow{3}{*}{anti-top}& \textit{total}     & $2.45$   & 
	$2.45$   &  $1.78$   &  
	$1.78$   &  $1.62$    &  
	$1.62$   \\
\cmidrule(l){2-8} 
	 & production      &    &    & $-0.46$    &  $-0.460(1)$   &  $-0.15$    &  
	 $-0.153(3)$    \\
\cmidrule(l){2-8} 
	& decay     &    &    &  $-0.21$   &  $-0.210(1)$   &  $-0.08$    &  $-0.078(1)$     \\ 
	\cmidrule(l){2-8} 
	& prod.~$\times$~decay     &    &    &  $$   &     &  $+0.07$    &  $+0.072(2)$     \\ 
	\bottomrule
\end{tabular}

	\egroup
	\label{table:BGZ-fiducial}
\end{table}

In contrast to our fiducial results that we present in the next section, the authors of 
ref.~\cite{Berger:2017zof} use \NNLO{} \PDF{}s throughout their study, and so also for \NLO{} predictions.
They furthermore estimate scale uncertainties using a two-point variation ($\mu_R=\mu_F=k\cdot m_t, 
k=\nicefrac{1}{2}, 2$) and a quadrature procedure from production and decay 
contributions. Their scale uncertainties of $^{+4.1\%}_{-2.2\%}$ and $^{+1.2\%}_{-0.7\%}$ at \NLO{}
and \NNLO{}, respectively, increase to $^{{+6.1\%}}_{{-4.0\%}}$ at \NLO{} and 
${}^{+2.3\%}_{-2.1\%}$ at \NNLO{} when a standard six-point scale variation (defined later, in \cref{sec:results})
is used.\footnote{When using \DDIS{} scales, the \NLO{} result and 
uncertainties change slightly to 
	$2.93^{+6.6\%}_{-6.1\%}$, but the \NNLO{} scale uncertainties increase somewhat more to 
	$2.67^{+5.7\%}_{-5.0\%}$.} These uncertainties are mostly driven by the renormalization 
	scale variation. 

The use of \NNLO{} \PDF{}s leads to \NNLO{} corrections of about $-9\%$ 
relative to \NLO{}, see \cref{table:BGZ-fiducial}, and the two-point quadrature scale-variation 
procedure leads to relatively 
small uncertainties, such that one observes a large gap between the \NLO{} and \NNLO{} predictions.
This large difference shrinks to about $-5\%$ when \NLO{} \PDF{}s are used consistently at 
\NLO{}, and one finds agreement within scale uncertainties. The consistent use of \PDF{}s 
is crucial for the $t$-channel single-top-quark process. In fact, the authors 
speculate about the smallness of the \emph{total inclusive} \NNLO{} corrections due to \DIS{} data 
used for \PDF{} fits and that the $t$-channel process mimics double deep inelastic scattering.
To fully exploit this property, it is essential that the order of the \PDF{}s is consistent with 
the hard scattering cross section order. This property transfers also to the fiducial region to 
some extent, but depends on the inclusiveness of the cuts. The double \DIS{} aspect can be 
maximally exploited in the factorized vertex-correction approach that we work with, namely by using 
\DDIS{} scales, where one then expects a maximum of perturbative stability of predictions.
This point will be discussed further in the following section.

\section{Results}
\label{sec:results}

In this section we examine both the fully inclusive cross sections and
the differential distributions relevant to experimental analyses.
Stability of inclusive cross sections between perturbative orders is a
signature of $t$-channel single-top-quark production, particularly when calculated
with the \DDIS{} scales, as it is effectively undoing the \DIS{} fits to the
\PDF{}s \cite{Sullivan:2017aiz}. 
The differential distributions are used
for a variety of experimental signatures as both a signal process and
as a background to any process that includes $W+$jets, e.g.\ $WH$ or
supersymmetry.  We describe below large differences between \NLO{} and \NNLO{}
in key distributions that are used to distinguish signal from background.

We begin by presenting fully inclusive results for \SI{7}{\TeV} and \SI{14}{\TeV} proton-proton collisions 
(\LHC{}) and for \SI{1.96}{\TeV} proton-antiproton collisions (Tevatron) in 
\cref{table:inclusive-ct14}. We show results using fixed scales $\mu_R=\mu_F=m_t$ for 
comparison with other results, and using the \DDIS{} scales.
Scale uncertainties are obtained using a six-point scale variation by evaluating 
the cross section for the scale choices $(k_F \cdot\mu_F; k_R \cdot\mu_R)$, where $\mu_F$ and 
$\mu_R$ are the central scales and
$$
(k_F;k_R) \in \{(2,2),(0.5,0.5),(2,1),(1,1),(0.5,1),(1,2),(1,0.5)\}\,.
$$
Subsequently maximum and minimum values are taken as the envelope.
Our default \PDF{} set in this section is \CTFOURTEEN{} \cite{CT14} and we use it at a consistent order together
with the partonic cross section order. That is, we use \NLO{} \PDF{}s for our \NLO{} prediction and 
\NNLO{} \PDF{}s for our \NNLO{} prediction.
\PDF{} uncertainties are given by 
evaluating the 56 eigenvector members (using \DDIS{} 
scales).\footnote{There are no such eigenvectors at \LO{} and we therefore do not show \PDF{} 
uncertainties at \LO{}.}

\begin{table}
	\centering
	\caption{Fully inclusive results in pb for $pp$ at \SI{7}{\TeV} and \SI{14}{\TeV} 
	(\LHC{}),
			as well as $p\bar{p}$ at \SI{1.96}{\TeV} (Tevatron) with scales $\mu_R=\mu_F=m_t$ and
			\DDIS{} scales and using {\abbrev CT14} \PDF{}s. Uncertainties next to the cross 
			section in super- and 
			subscript are from a six-point
			scale variation, while \PDF{} uncertainties are below.}
	\vspace*{0.5em}
	\bgroup
	\setlength\tabcolsep{0.5em}
	\def\arraystretch{1.5}%

\begin{tabular}{@{}l|l|l|l|l|l@{}}

	\toprule

	&  \multicolumn{2}{c|}{\SI{7}{\TeV} $pp$} & 
	\multicolumn{2}{c|}{\SI{14}{\TeV} $pp$}  & \SI{1.96}{\TeV} $\bar{p}p$   \\ \midrule
	& top    & anti-top & top      & anti-top & $t+\bar{t}$ \\
\midrule
	
	$\sigma_\text{LO}^{\mu=m_t}$   &    $37.1^{+7.1\%}_{-9.5\%}$      & $19.1^{+7.3\%}_{-9.7\%}$  
	&  $134.6^{+10.0\%}_{-12.1\%}$   & $78.9^{+10.4\%}_{-12.6\%}$   & $2.09^{+0.8\%}_{-3.1\%}$    \\

	$\sigma_\text{LO}^\DDIS{}$ & $39.5^{+6.4\%}_{-8.6\%}$  & $19.9^{+7.0\%}_{-9.3\%}$    &   
	$140.9^{+9.4\%}_{-11.4\%}$     & $80.7^{+10.2\%}_{-12.3\%}$    &  $2.31^{-0.3\%}_{-1.8\%}$     
	\\ \midrule
		$\sigma_\text{NLO}^{\mu=m_t}$   & $41.4^{+3.0\%}_{-2.0\%}$      & 
		$21.5^{+3.1\%}_{-2.0\%}$     & $154.3^{+3.1\%}_{-2.3\%}$   & $91.4^{+3.1\%}_{-2.2\%}$    &  
		$1.96^{+3.1\%}_{-2.3\%}$  \\

	$\sigma_\text{NLO}^\DDIS{}$ &  $41.8^{+3.3\%}_{-2.0\%} $     & 
	$21.5^{+3.4\%}_{-1.6\%}$     & 
	$154.4^{+3.7\%}_{-1.4\%}$       &   $91.2^{+3.1\%}_{-1.8\%}$   &  
	$2.00^{+3.6\%}_{-3.4\%} $      \\

	& \PDF{}$\,^{+1.7\%}_{-1.4\%}$ & \PDF{}$\,^{+2.2\%}_{-1.5\%}$ & \PDF{}$\,^{+1.7\%}_{-1.1\%}$ & 
	\PDF{}$\,^{+1.9\%}_{-0.9\%}$ 
	& \PDF{}$\,^{+4.3\%}_{-5.3\%}$
	\\
	 \midrule
	
		$\sigma_\text{NNLO}^{\mu=m_t}$   & $41.9^{+1.2\%}_{-0.7\%}$  & $21.9^{+1.2\%}_{-0.7\%}$   
		& $153.3(2)^{+1.0\%}_{-0.6\%}$   & $91.5(2)^{+1.1\%}_{-0.9\%}$  &  
		$2.08^{+2.0\%}_{-1.3\%}$
		  \\
	$\sigma_\text{NNLO}^\DDIS{}$ &  $41.9^{+1.3\%}_{-0.8\%}$     &  
	$21.8^{+1.3\%}_{-0.7\%}$    &  $153.4(2)^{+1.1\%}_{-0.7\%}$      &  
	$91.2(2)^{+1.1\%}_{-0.9\%}$     &   
	$2.07^{+1.7\%}_{-1.1\%}$   \\
	 & \PDF{}$\,^{+1.3\%}_{-1.1\%}$ & \PDF{}$\,^{+1.4\%}_{-1.3\%}$  & 
	 \PDF{}$\,^{+1.2\%}_{-1.0\%}$ 
	 & 
	 \PDF{}$\,^{+1.0\%}_{-1.0\%}$ & \PDF{}$\,^{+3.7\%}_{-5.0\%}$
	\\
	 \bottomrule

\end{tabular}
	\egroup
	\label{table:inclusive-ct14}
\end{table}

While there are noticeable differences in the inclusive cross section
between \DDIS{} scales and $\mu=m_t$ at \LO{}, the differences are
within scale uncertainties at the LHC.  At the Tevatron the large
difference is due to cancellations between corrections on the
light-quark line and the bottom-quark line, and the \LO{} scale uncertainties
grow to $\pm 7\%$ when varied independently \cite{Harris:2002md}.
Overall we see that the \NNLO{} corrections are almost zero and
that \NLO{} and \NNLO{} results overlap well within the small scale
uncertainties of $\sim1\%$. \PDF{} uncertainties are at a similar
level, but larger for the Tevatron.

For the Tevatron it is noteworthy that when using the \DDIS{} scales and \CTFOURTEEN{}
\PDF{}s the cross section varies by 10--15\% between \LO{} and
\NLO{} \cite{Sullivan:2017aiz}.  This large difference, which should vanish for pure \DIS{} data
and fully correlated fits, shrinks to a few percent between \NLO{}
and \NNLO{}. Whether this is an artifact of these particular \PDF{} fits or is a sign of a
deeper problem with the way \PDF{}s are parameterized is material for a
dedicated study \cite{CNZprod}. 

\subsection{Fiducial and differential cross sections}
\label{sec:fiducial}

The key to understanding signals and backgrounds in 
$t$-channel single-top-quark analyses are the $W$+$n$-jet exclusive cross sections. The
$2$-jet cross section with exactly one $b$-quark tag is enriched with the $t$-channel signal, while 
the $3$-jet 
cross section with one or two $b$-quark tags is dominated by background processes like $t\bar{t}$. It 
is essential that precise predictions are available for both the signal region of $t$-channel 
single-top-quark production, but also for the
background region, which subsequently constraints the 
$t\bar{t}$ contribution in final fits. In practice, the final extraction is done
using a multivariate analysis with various discriminator observables, see e.g. 
ref.~\cite{Sirunyan:2016cdg}.

Our initial acceptance cuts are inclusive and defined
in \cref{tab:fiducialcuts}.  We require \emph{at least} one $b$-tagged
jet and one non-$b$-tagged (light) jet in order to study the effect of
the \NNLO{} corrections on jet counting.  In the results presented
below we reconstruct the top-quark momentum by adding the exact $W$-boson
momentum and the momentum of the $b$-jet with the
largest $p_T$.  We are particularly interested in reconstructed $t$+$n$-jet 
observables.  In the following $j_b$ denotes a $b$-tagged jet and $j_l$ a
light-quark jet (not $b$ tagged), while $j$ without any subscripts labels
any kind of jet.

\begin{table}
	\centering
	\caption{Our fiducial cuts at $\sqrt{s}=\SI{13}{\TeV}$.}
	\vspace*{0.5em}
	\bgroup
	\setlength\tabcolsep{1em}
	\def\arraystretch{1.5}%
	\begin{tabular}{l|c}
		Lepton cuts & $p_T^{l} > \SI{25}{\GeV}, \abs{\eta^l} < 2.5$\\
		anti-$k_T$ jet clustering & $p_T^{j} > \SI{30}{\GeV}, R=0.4, \abs{\eta_j}<4.5$\\
		jet requirements &	at least one non-$b$ (light) jet and at least one $b$ jet
	\end{tabular}
	\egroup
	\label{tab:fiducialcuts}
\end{table}

For our set of cuts we find that, when \PDF{}s are evaluated at the same order as the matrix element correction, the total 
fiducial \NNLO{} prediction $\sigma^\NNLO{}=\SI{5.75}{pb}$ agrees with the \NLO{} prediction 
$\sigma^\NLO{}=\SI{5.65}{pb}$ within less than two percent (see \cref{tab:ourfiducial_jetcross}). 
The stability across orders is due to a consistent use of \PDF{}
orders.\footnote{Generally, if \NNLO{} \PDF{}s were used with \NLO{} matrix
elements, the difference would increase by a few percent.}  
We also see a significant reduction in the scale uncertainty of the
individual exclusive $tj$ and $tjj$ channels from $\sim 10$--$15\%$ at \NLO{} to
$\sim 5$--$8$\% at \NNLO{}.  The prediction for $tjjj$ is a \LO{} prediction
with correspondingly larger uncertainties.  The $tjjj$ exclusive channel
uncertainty dominates the semi-inclusive $t j_l$ fiducial uncertainty at \NNLO{}, which is
therefore not much smaller than the \NLO{} uncertainty.

However, while the semi-inclusive cross section is stable with small
uncertainties, there is a large 20\% shift of events from the
$t+2$-jet bin to the $t+3$-jet bin (along with a small reduction of
the cross section in the $t+1$-jet signal bin) when moving from \NLO{}
to \NNLO{}.  This is very significant because the $t+2$-jet bin is
precisely the bin on which the cut is made to separate signal from
background.  In practice, the experimental data in each of the jet bins is
normalized to the relative fraction predicted by theory, and the
absolute cross section is floated to match the inclusive data.  This
means that the while the signal prediction in $tj$ goes down by 5\%
at \NNLO{}, too many events are being cut if the \NLO{} normalization
is used for the $tjj$ bin.  Hence, the net signal predicted by \NNLO{}
will increase compared to \NLO{}, and the $tjj$ background to other
physics processes is smaller than expected by \NLO{}.

\begin{table}
	\centering
	\caption{Cross sections for the production of a top quark ($W$ plus $p_T$-leading $b$-jet) and 
	additional 
	jets using \DDIS{} scales, in pb. The cross sections 
	are given in the fiducial region 
	as in \cref{tab:fiducialcuts}, but apart from the inclusive $t j_l$ row the jet 
	requirements are adjusted accordingly. $j_b$ denotes a $b$-tagged jet 
	and $j_l$ a light-quark jet (not $b$ tagged). Contributions with more than one additional 
	$b$-tagged jet are negligible and omitted. Uncertainties from a six point scale variation 
	are given in super- and subscript. Percentages in parenthesis give the fraction with respect to 
	the individual inclusive jet category. Numbers in bold font in each column add up to the $tj_l$ 
	inclusive result within numerical uncertainties.}
	\vspace*{0.5em}
	\bgroup
	\setlength\tabcolsep{1em}
	\def\arraystretch{1.5}%
\begin{tabular}{@{}c|cccc@{}}

	\toprule

	& \LO{} & \NLO{} & \NNLO{} &  \\ \midrule

	$tj_l$ inclusive  &  $5.51^{+9.1\%}_{-11\%}$   & ${5.65}^{+3.8\%}_{-3.1\%}$  &  
	${5.75}^{+3.3\%}_{-2.7\%}$ &  \\

	\midrule
	$tj$ &  $5.51^{+9.1\%}_{-11\%}$   & $3.77^{+10.0\%}_{-9.8\%}$ (100\%)  & 
	$3.55^{+7.1\%}_{-5.9\%}$ (100\%)  &   
	\\

	$tj_l$ & $5.51^{+9.1\%}_{-11\%}$    &  $\mathbf{3.24}^{+13.6\%}_{-13.2\%}$ (86\%) &  
	$\mathbf{3.08}^{+8.2\%}_{-6.5\%}$ (87\%)
	&   \\
	$tj_b$ & ---  & $0.53^{+14.9\%}_{-12.2\%}$ (14\%) & $0.47^{+2.5\%}_{-5.3\%}$ (13\%)&\\ \midrule
	$tjj$ & --- & $\mathbf{2.42}^{+10.4\%}_{-9.2\%}$ (100\%) & $\mathbf{1.90}^{+4.3\%}_{-7.6\%}$ 
	(100\%) &\\
	$tj_l j_l$ & --- & $0.93^{+10.6\%}_{-8.7\%}$ (38\%) & $0.75^{+4.2\%}_{-5.9\%}$ (40\%) &\\
	$t j_b j_l$ & --- & $1.49^{+15.3\%}_{-12.5\%}$ (62\%) & $1.15^{+4.5\%}_{-9.9\%}$ (60\%) &\\ 
	\midrule
	$t j j j$ & --- & --- & $\mathbf{0.77}^{+27.3\%}_{-20.0\%}$ (100\%) & \\
	$t j_l j_l j_l$ & --- & --- & $0.13^{+20.0\%}_{-15.1\%}$ (17\%) &\\
	$t j_b j_l j_l$ & --- & --- & $0.64^{+30.0\%}_{-21.4\%}$ (83\%) &\\
	 \bottomrule

\end{tabular}
	
	\egroup
	\label{tab:ourfiducial_jetcross}
\end{table}

\paragraph{Differential cross sections.}
We now move on to discuss differential distributions. For all the following plots we show absolute 
distributions with scale uncertainties in the upper 
panel, while in the lower panel we show the ratios $\NNLO{}/\NLO{}$ and $\NLO{}/\NLO{}$ to study 
perturbative stability across orders. For the ratios we show the scale uncertainties from variation
in the numerator only. Perturbative stability across 
orders can therefore be judged by examining the extent to which the uncertainty bands overlap.

\begin{figure}
	\centering
	\includegraphics{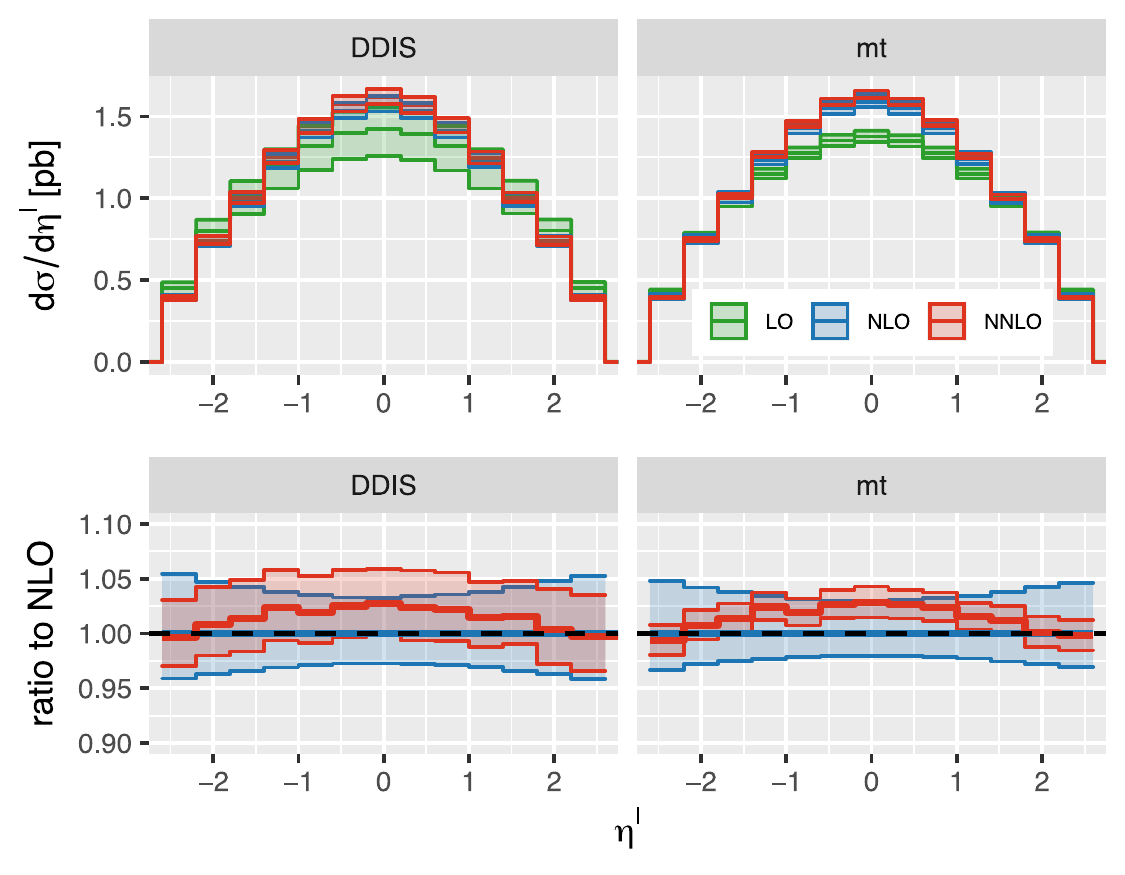}
	\caption{Pseudorapidity distribution of the charged lepton with \DDIS{} scales (left) and fixed 
		scale $m_t$ (right).}
	\label{fig:fiducial_etal}
\end{figure}

We begin our discussion with the charged lepton pseudorapidity
distribution shown in \cref{fig:fiducial_etal} at \LO{}, \NLO{} and \NNLO
{}.  This distribution is mostly kinematically driven and, as expected, 
the \NNLO{} corrections are consistent with zero to within a few percent fully 
differentially. The differences
in the central prediction between \DDIS{} scales and fixed scale $m_t$ are small at \NNLO{}, but we 
observe noticeably larger scale uncertainties using the \DDIS{} scales, allowing for a robust 
overlap of predictions from \LO{} through \NNLO{}. The relatively large scale uncertainties
at \NNLO{}, compared to \NLO{}, were already visible in the $tj_l$ inclusive cross section in 
\cref{tab:ourfiducial_jetcross} and are an artifact of
the specific set of rather inclusive cuts. Compare this, for example, with the \emph{fully} 
inclusive uncertainties in \cref{table:inclusive-ct14}, which are much smaller at the level of 
$1\%$ and the large uncertainty decrease in the individual jet cross sections in 
\cref{tab:ourfiducial_jetcross}. Overall, the large scale uncertainties are also consistent with 
the 
observation that the \NNLO{} effects are almost as large as the \NLO{} effects.

For all other distributions we find 
similar central results between \DDIS{} and $m_t$ scales.  Since \DDIS{} scales are the most 
consistent 
with the DIS nature of the $t$-channel process we only show the \DDIS{} results with their more 
robustly estimated scale uncertainties.

\begin{figure}
	\centering
	\includegraphics{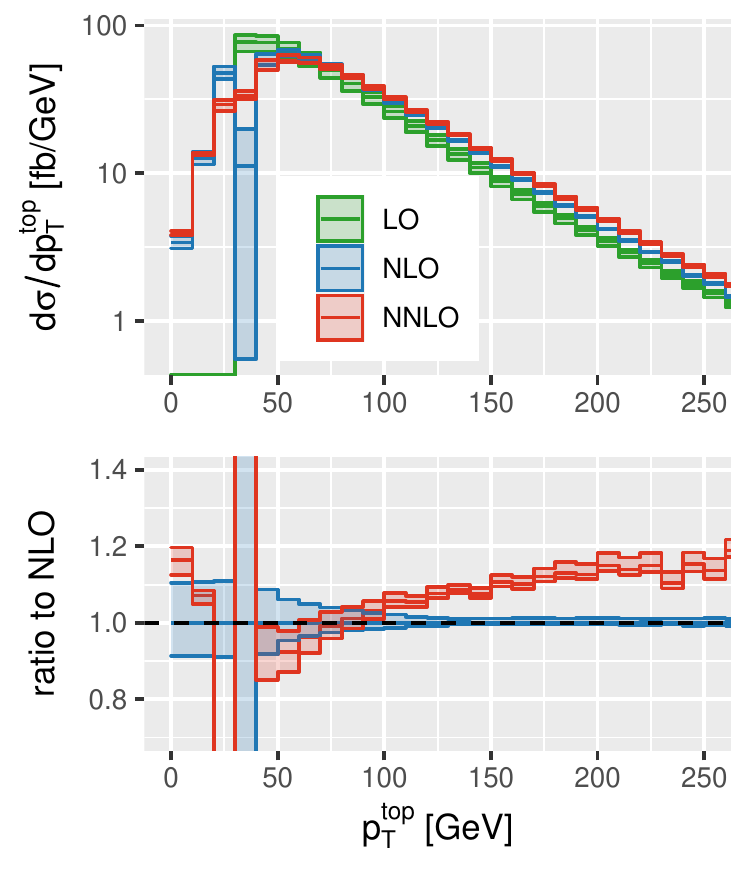}
	\includegraphics{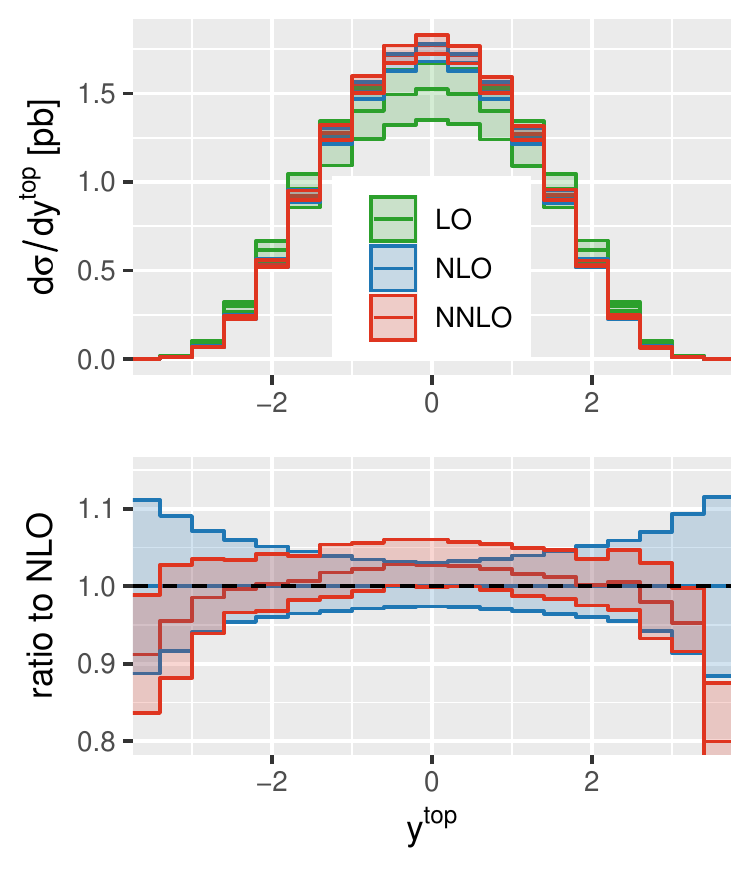}
	\caption{Top-quark transverse momentum distribution (left) and rapidity distribution (right).}
	\label{fig:toppt}
\end{figure}

Other standard observables to consider are the top-quark transverse momentum and rapidity. We 
show these distributions in \cref{fig:toppt}. The \NNLO{} rapidity corrections are small and 
consistent with zero within scale uncertainties. The \NNLO{} transverse momentum corrections are, 
on the other hand, sizable and correct \NLO{} results by up to +15\% at \SIrange{200}{250}{\GeV}.
Kinematically, the top quark $p_T$ is driven by recoil from additional jets beginning at an \NLO{} 
calculation. Therefore, one expects to see large effects at \NNLO{}, which we observe here
and which are outside scale uncertainty bands. The kinematic jet recoil threshold around 
$\SI{30}{\GeV}$ is also significantly stabilized at \NNLO{} by the additional radiation. This 
threshold region had to be addressed previously by a parton shower or partial resummation.

\begin{figure}
	\centering
	\includegraphics{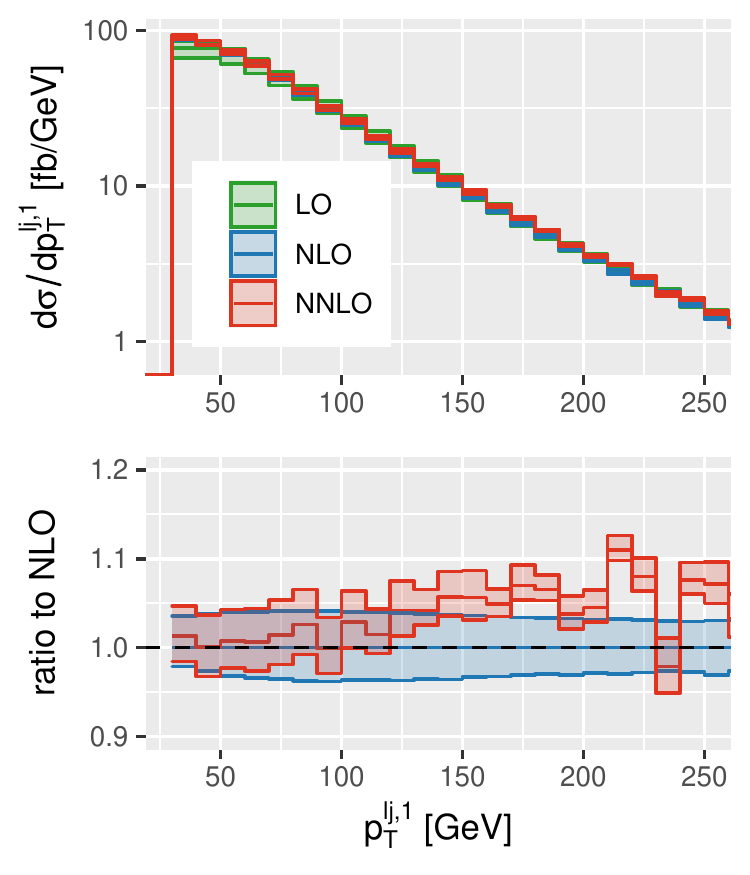}
	\includegraphics{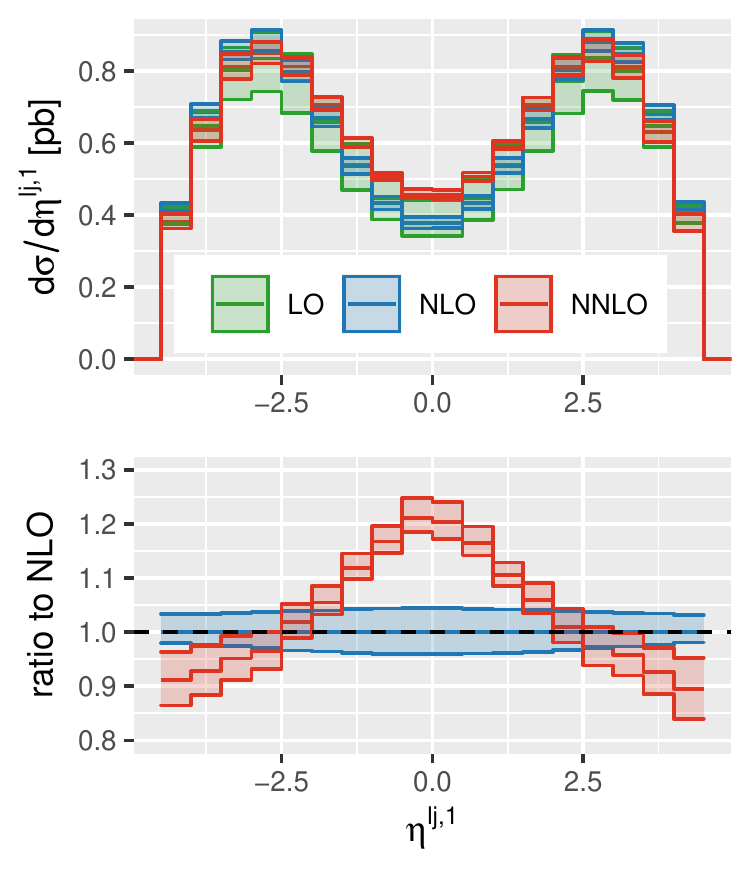}
	\caption{$p_T$-leading light jet transverse momentum distribution (left) and pseudorapidity 
		distribution (right).}
	\label{fig:lj1}
\end{figure}

\paragraph{$t$-channel signal and background.}
In addition to jet counting, 
the most important discriminatory observable for the signal is the light-quark jet 
pseudorapidity, which in 
$t$-channel production has its distinctive peak in the forward direction. 
In 
\cref{fig:lj1} we present the ($p_T$) leading light-jet transverse momentum and 
pseudorapidity distributions.
\NNLO{} corrections drive the transverse momentum slightly harder in 
the tail but have little effect in the peak region. The shape of the pseudorapidity distribution 
changes noticeably, with negative corrections of $10\%$ in the very forward region and positive 
corrections of $20\%$ in the central region, leaving the peak region with corrections of just a 
few percent.

\begin{figure}
	\centering
	\includegraphics{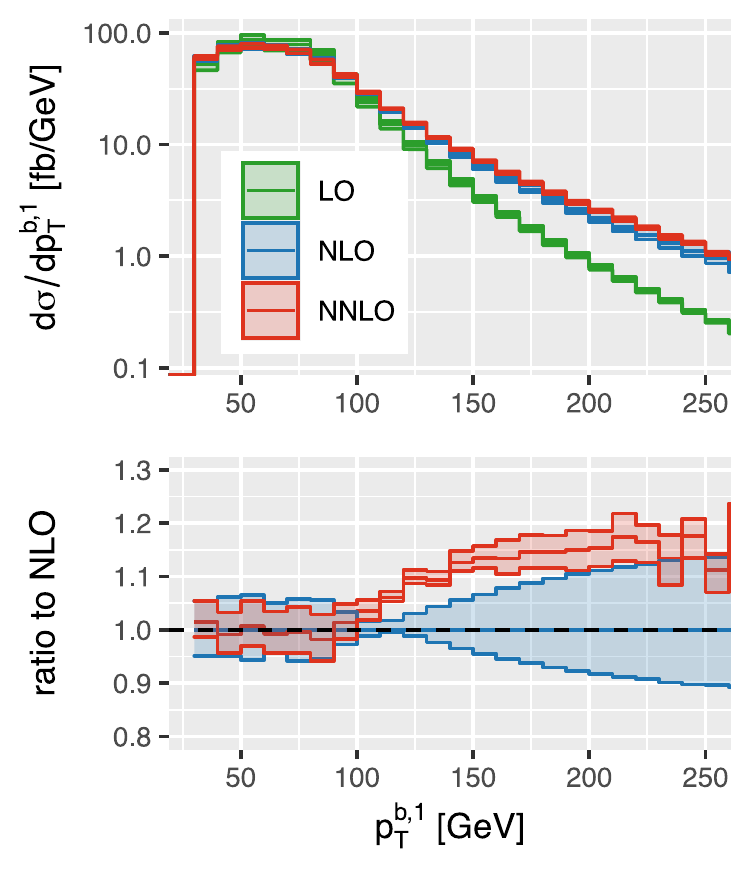}
	\includegraphics{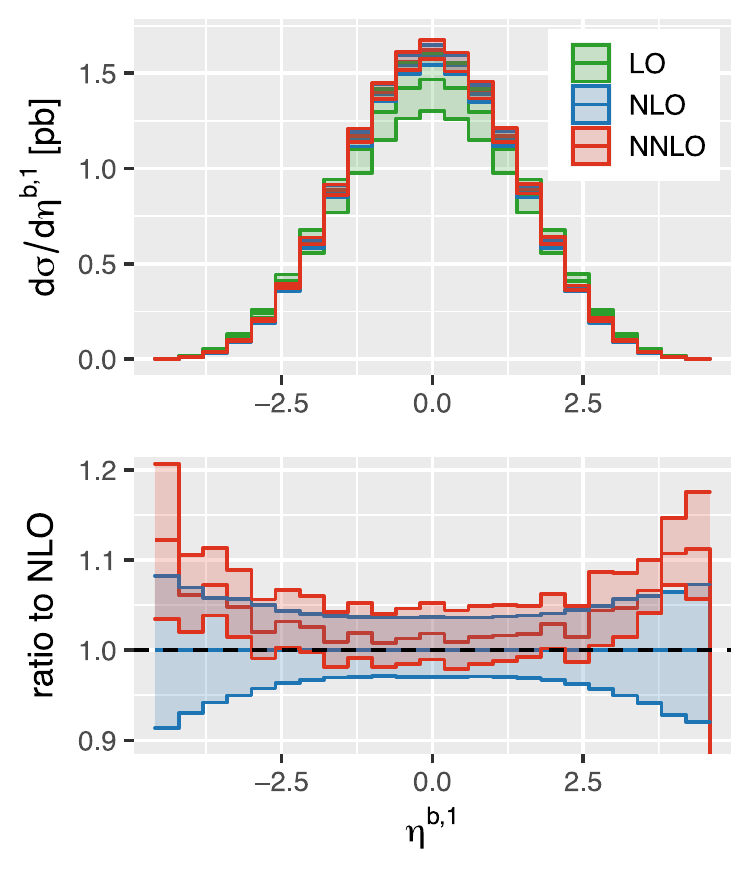}
	\caption{$p_T$-leading $b$-tagged jet transverse momentum distribution (left) and 
		pseudorapidity 
		distribution (right).}
	\label{fig:b1}
\end{figure}

The leading $b$-jet distributions shown in \cref{fig:b1} are relevant for the top-quark 
reconstruction and enter directly in our previously shown top-quark distributions. For the 
transverse momentum we observe zero corrections in the peak region but positive corrections of 
$15\%$ at large $p_T$ driven by additional recoil available at \NNLO{}. The pseudorapidity 
distribution receives only positive corrections of $2-3\%$ in the relevant peak region.

Another important discriminatory observable is the angle between the lepton and the leading
light-quark jet in the top-quark rest frame, which we discuss later.

\begin{figure}
	\centering
	\includegraphics{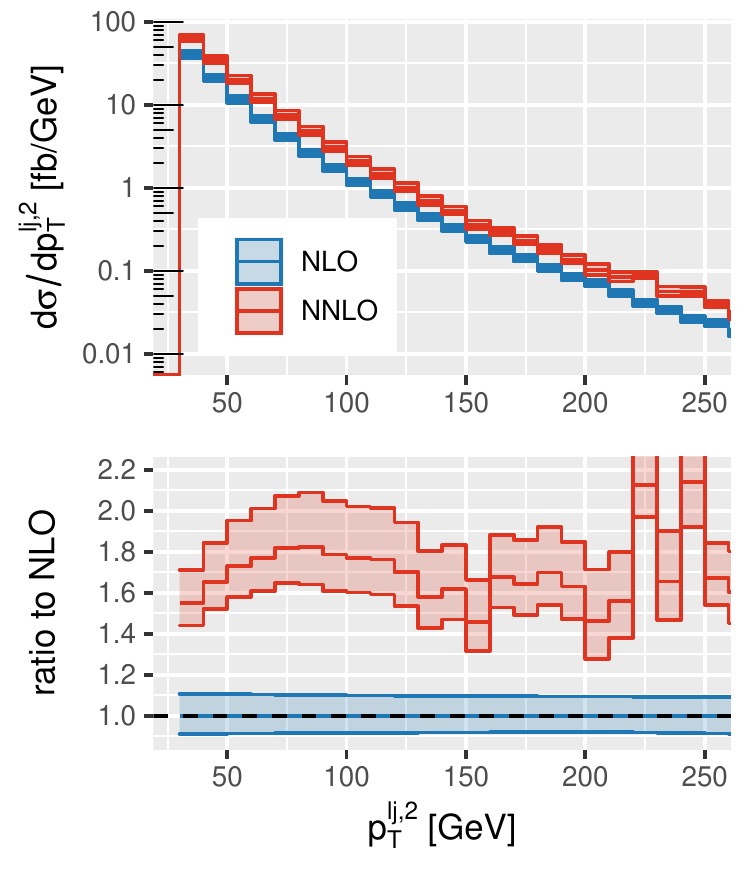}
	\includegraphics{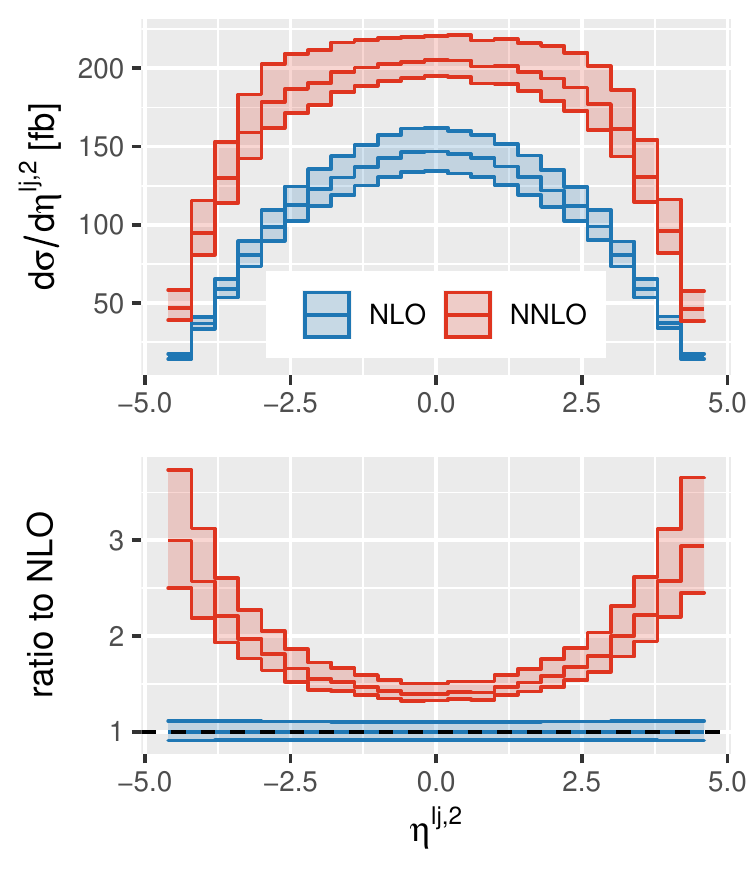}
	\caption{$p_T$-subleading light jet transverse momentum distribution (left) and pseudorapidity 
		distribution (right).}
	\label{fig:lj2}
\end{figure}

\begin{figure}
	\centering
	\includegraphics{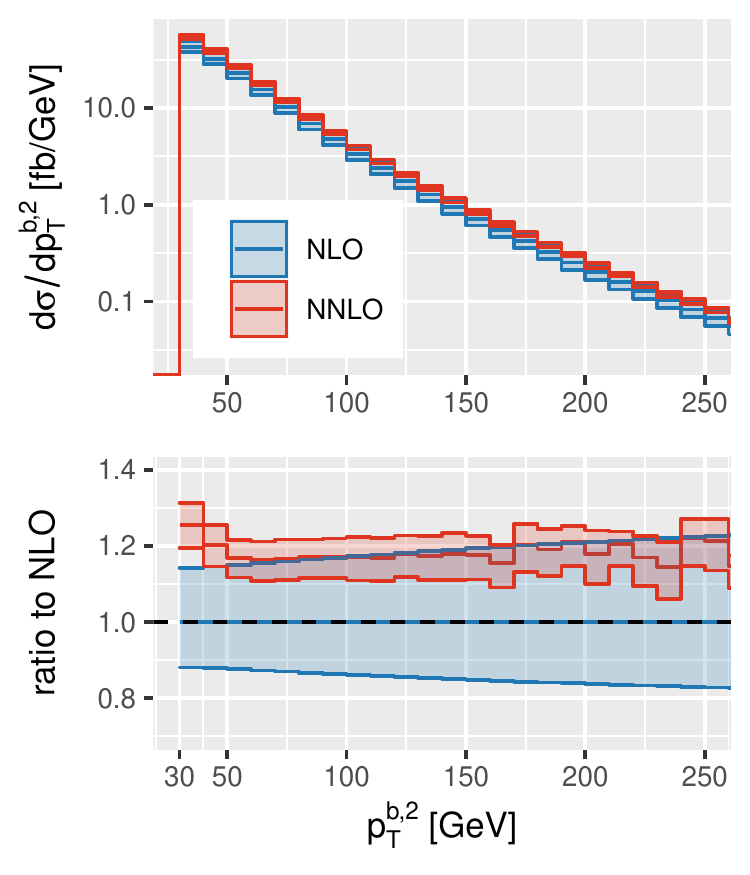}
	\includegraphics{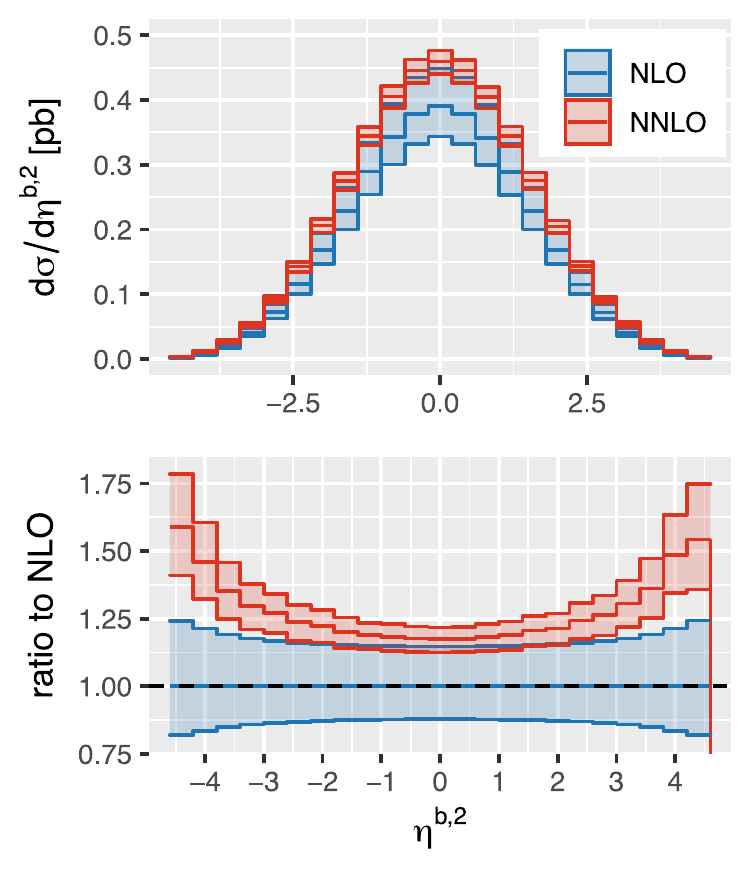}
	\caption{$p_T$-subleading $b$-tagged jet transverse momentum distribution (left) and 
		pseudorapidity 
		distribution (right).}
	\label{fig:b2}
\end{figure}

Moving to the relevant distributions for a proper background estimation, we show the subleading 
light-quark jet and subleading $b$-tagged jet transverse momentum and pseudorapidity distributions 
in \cref{fig:lj2,fig:b2}, respectively. Since these distributions enter for the first time
at \NLO{} in our five-flavor scheme calculation, the $\alpha_s$ corrections at \NNLO{} are large 
and significant. And while these distributions could be obtained from an \NLO{} event generator
with a Born process of single-top-quark production plus an additional jet ($W$+3-jet production),
the calculation within our \NNLO{} framework allows for a correct normalization of these background
contributions to the signal contribution.

In \cref{sec:additional_figures} we present additional distributions for the lepton transverse 
momentum, see \cref{fig:ptl_etal}, and the $b$-quark-lepton system invariant mass and transverse 
momentum, see \cref{fig:ptbl_mbl}.

\begin{figure}
	\centering
	\includegraphics{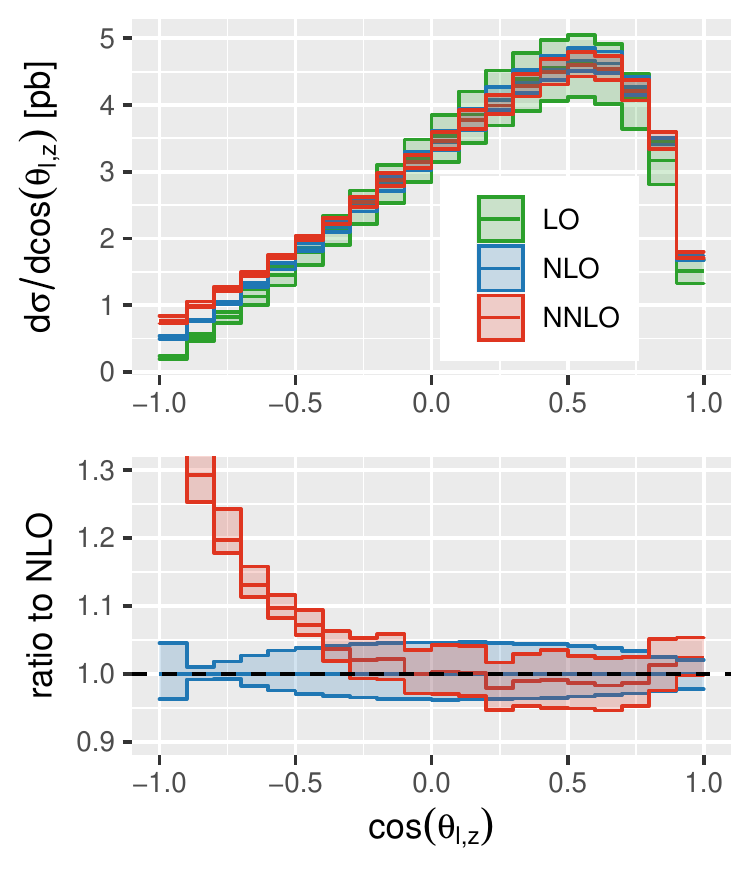}
	\caption{Angular distribution for $\coszprod$.}
	\label{fig:coszprod}.
\end{figure}

\paragraph{Angular observables in the top-quark rest frame.}
In addition to assessing the impact of higher order corrections on the
usual kinematical distributions such as invariant masses, transverse momenta
and rapidities, it is important to also consider their effect on angular correlations.  The angle
between the leading non-$b$ jet and the lepton from the top-quark
decay in the top-quark rest frame \cite{Sullivan:2005ar} exhibits a strong 
correlation~\cite{Mahlon:1996pn,Mahlon:1999gz} and is one of the key observables used to identify
the $t$-channel process. Our \NNLO{} results for this observable $\cos(\theta_{l,z})$ are shown in 
\cref{fig:coszprod} and we find that \NNLO{} corrections are consistent with zero for the bulk 
region. At large angles ($\cos\theta_{l,z} \sim -1$) the additional radiation at \NNLO{} becomes
important and the corrections are significant, as expected.

Several other angular distributions measured in the top-quark or $W$-boson rest frame are expected 
to be modified by non-standard model
physics effects~\cite{Aguilar-Saavedra:2014eqa,AguilarSaavedra:2010nx}. Such angles are constructed 
to be
sensitive to new physics in the production and decay stages of the top quark, respectively. We 
present \NNLO{} results for two such sets of angles in \cref{sec:appendix_angles} and find
that \NNLO{} corrections are mostly consistent with zero in the bulk regions within a few percent 
of scale uncertainties. For the study of anomalous 
couplings and \SMEFT{} contributions this perturbative stability at \NNLO{} is important, since 
such analyses are performed at \NLO{}. In fact, the inclusion
of off-shell effects at \NLO{} has previously been found to be equally or more important for some 
angular distributions \cite{Neumann:2019kvk}.

\section{Conclusions}

We have presented a calculation of $2\to4$ (five-flavor scheme) $t$-channel single-top-quark 
production 
and decay at \NNLO{}.  Due to the special importance of the top quark in the Standard Model, and the
availability of large, high-quality data sets from the LHC experiments, predictions
at this level of sophistication are mandatory for future analyses.  Moreover,
such calculations should be cross-checked carefully and made available publicly in order to facilitate
any such studies.

 Our calculation is performed in the on-shell and vertex-function 
approximation that allows for a factorization into \NNLO{} corrections in production on the 
light-quark line, heavy-quark line and corrections in decay. We include all 
\NNLO{} 
vertex corrections through three \NNLO{} calculations but also the $\NLO{}\otimes\NLO{}$ 
interference contributions between 
light line 
and heavy line, as well as production 
and decay at the amplitude level, thereby preserving all spin correlations between production and 
decay.
We performed extensive checks of all components of our calculation, and at all stages of the assembly. This 
includes analytical checks of all ingredients entering the factorization theorems used for our 
\NNLO{} subtractions,  comparisons of individual amplitudes with the numerical one-loop library Recola, and 
verification of our numerical implementations of dipole subtractions and slicing subtractions at the 
per-mille level.

With the results of our calculation we have been able to scrutinize
the results of the stable top-quark calculation \cite{Brucherseifer:2014ama} and 
the results with decay in refs.~\cite{Berger:2016oht,Berger:2017zof}. We find full agreement 
with the latter calculation at the per-mille level, but in the former case find differences of 
100\% on the total \NNLO{} \emph{coefficient} and up to six percent in the \emph{full} \NNLO{} 
top-quark $p_T$ 
distribution. Our calculation therefore resolves this discrepancy and, furthermore, validates the 
first implementation of differential \NNLO{} corrections in the decay 
\cite{Gao:2012ja,Berger:2016oht,Berger:2017zof}.

A special focus of our implementation is to keep full flexibility for dynamic factorization and 
renormalization scales in all parts of the calculation. One is then able to exploit that 
single-top-quark production is like double deep inelastic scattering (\DDIS{}) in the approximation 
that we use, and set the scales as they are used in the fitting of \PDF{}s from light-quark and 
heavy-quark \DIS{} data. While the effect of using \DDIS{} scales compared to using a fixed scale 
$m_t$ seems small at \NLO{} and \NNLO{} for the \PDF{} set considered here, it can serve as a 
consistency check and constraint of \PDF{}s that we intend to pursue in a subsequent study.

We presented total cross sections at the fully inclusive level including scale 
uncertainties and \PDF{} uncertainties for the \SI{7}{\TeV} and \SI{14}{\TeV} \LHC{} and for 
proton-antiproton collisions at the Tevatron for both a fixed central scale as well as \DDIS{} scales.
We studied fiducial cross sections at the total and the differential level in detail, presenting 
standard kinematical distributions used for single-top-quark analyses. We estimate scale 
uncertainties using a standard six-point variation and \DDIS{} central scales. With this 
prescription we find that uncertainties are about a factor of two larger than in the previous 
calculation \cite{Berger:2017zof}, which employed a different prescription, but
our choice better represents the differential variation order-by-order.

Overall we find that \NNLO{} corrections are crucial for a precision identification of the 
$t$-channel process, whose primary discriminatory observable is the leading light-quark jet 
pseudorapidity. This observable receives shape corrections of up to 20\% at \NNLO{}. Higher-order
effects are also important for the background categories of $W$+3-jet production where corrections 
are 50--100\%. We also showed \NNLO{} 
predictions for certain sets of angles defined in the top-quark rest frame that are sensitive to 
new physics in the production and decay vertices. We find that \NNLO{} corrections for these angles 
are small and that therefore these angles are kinematically robust and can be taken into account at 
\NLO{} at the current level of precision for the search for new physics.

In the future it could be interesting to investigate if the use of \DDIS{} scales can provide a 
useful constraint on \PDF{}s. The general relevance of $t$-channel single-top-quark production for 
\PDF{} fits has 
been studied before in ref.~\cite{Nocera:2019wyk}. To further improve the accuracy of our
predictions, \NLO{} off-shell effects implemented in \MCFM{} \cite{Neumann:2019kvk} could be 
incorporated with a reweighting procedure. Additionally, one could furthermore work towards the 
inclusion of
the $1/N_c^2$ color-suppressed effects from two-loop box diagrams, which should be feasible with 
modern loop-calculation techniques.

Our calculation will be made publicly available in an upcoming version of \MCFM{} to be useful 
directly for the \LHC{} precision physics program.

\paragraph{Acknowledgments}

We would like to thank Jun Gao for providing information regarding the
calculation presented in ref.~\cite{Berger:2017zof}:
for supplying us with more detailed cross section numbers for comparison,
that allowed for the exhaustive cross-check detailed in \cref{table:BGZ-fiducial}
and the surrounding discussion,
as well as for clarifying the details of the scale variation procedure.
We furthermore thank Jean-Nicolas Lang for providing us with Recola 4-flavor and 5-flavor scheme 
model files.

This document was prepared using the resources of
the Fermi National Accelerator Laboratory (Fermilab), a
U.S. Department of Energy, Office of Science, HEP User
Facility. Fermilab is managed by Fermi Research Alliance, LLC (FRA),
acting under Contract No.\ DE-AC02-07CH11359. Tobias Neumann is supported by the United States 
Department of Energy under Grant Contract DE-SC0012704.
The numerical calculations reported in this paper were performed using the Wilson High-Performance 
Computing Facility at Fermilab.

\appendix

\section{Additional fiducial distributions}
\label{sec:additional_figures}

In this Appendix we provide supplementary figures for select kinematic 
distributions and angular correlations using the fiducial cuts of \cref{tab:fiducialcuts} in 
\cref{sec:fiducial}.

\begin{figure}
	\centering
	\includegraphics{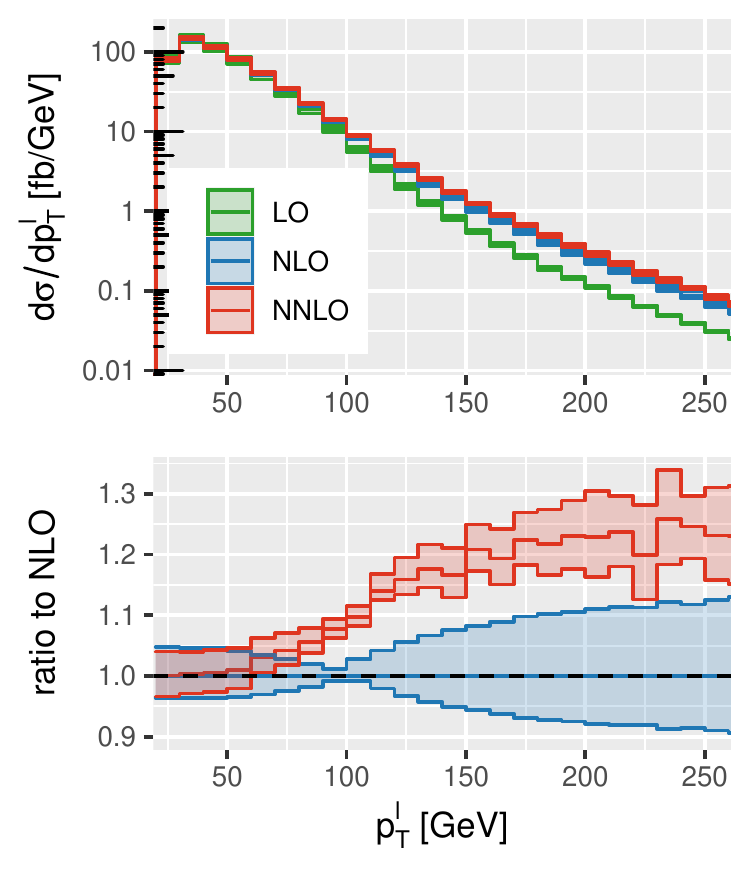}
	\includegraphics{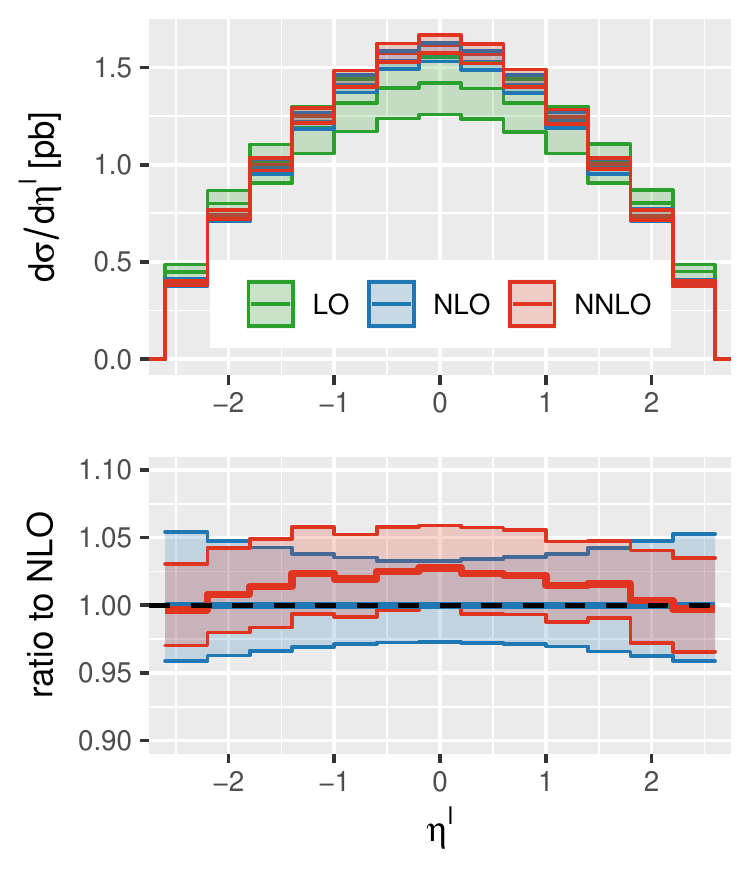}
	\caption{Positron transverse momentum distribution (left) and 
		pseudorapidity distribution (right).}
	\label{fig:ptl_etal}
\end{figure}

In \cref{fig:ptl_etal} we present the \NNLO{} transverse momentum and pseudorapidity 
distributions for the lepton from the top-quark decay. The \NNLO{} corrections in the 
pseudorapidity distribution are
at the level of a few percent but still consistent with zero within scale uncertainties,
as may be expected from our total fiducial cross section in \cref{tab:ourfiducial_jetcross}.
The transverse momentum distribution on the other hand receives sizable corrections in the
tail, but is perturbatively stable in the peak region. The corrections in the tail are large
because the transverse recoil is driven by the subleading jet in the decay, entering for the first 
time at \NLO{}.

\begin{figure}
	\centering
	\includegraphics{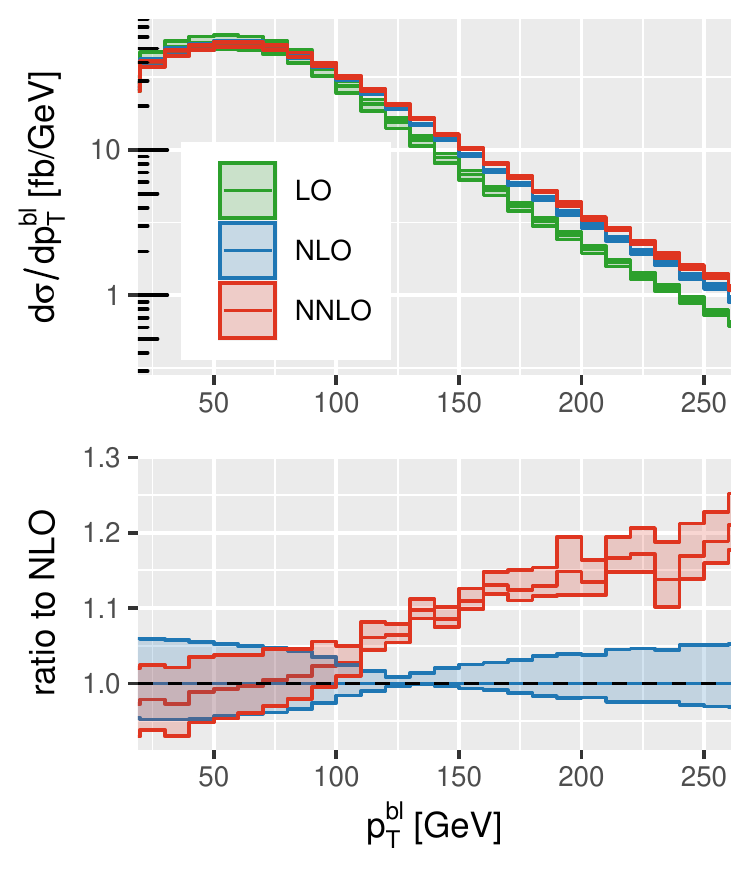}
	\includegraphics{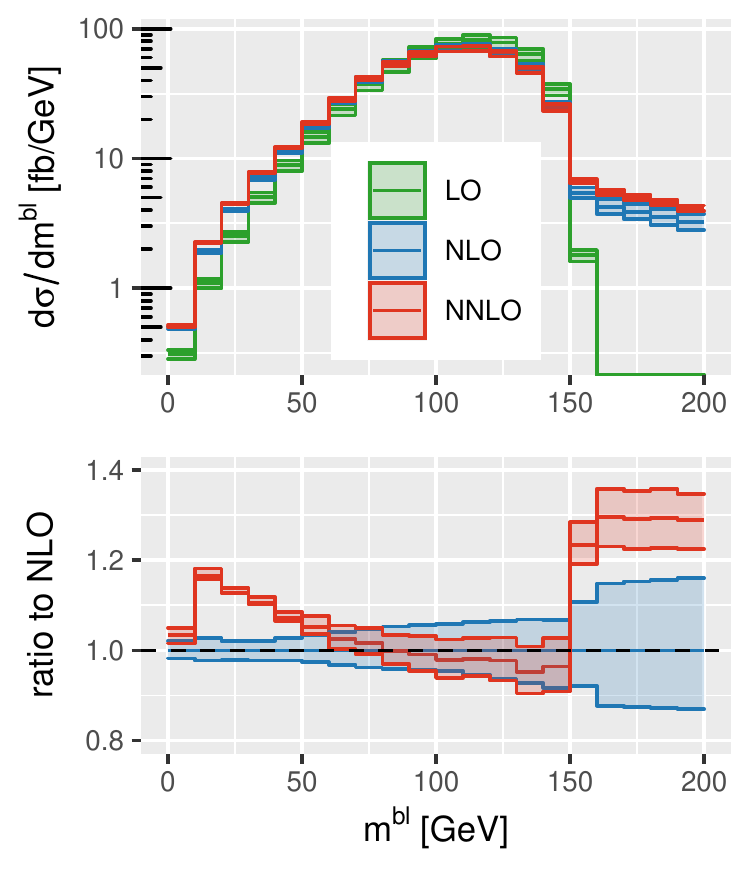}
	\caption{Leading $b$-tagged jet plus lepton transverse momentum distribution (left) and 
		invariant mass distribution (right).}
	\label{fig:ptbl_mbl}
\end{figure}

Combinations of the leading $b$-jet and lepton system are relevant for the top-quark mass 
measurement in $t$-channel production \cite{ATLAS:2014baa,Sirunyan:2017huu} since they 
allow one to circumvent the neutrino reconstruction. We show the $bl$ transverse momentum and invariant 
mass distribution in \cref{fig:ptbl_mbl} and find that while \NNLO{} corrections are small in the 
peak regions, they are important otherwise. 
The large perturbative corrections to the transverse momentum distribution
follow directly from the stiffening of the lepton transverse momentum at \NNLO{} that is included
in this variable.

\label{sec:appendix_angles}

We also present \NNLO{} results for two sets of angular observables that are
sensitive to modifications from new physics in the top-quark production
and decay stages, respectively.
To describe the first set of angles we introduce a coordinate system 
that uses the direction of the leading light-quark jet $\vec{s}_t$
in the top-quark rest frame to define the $z$-axis, $\hat{z}$. The leading light-quark jet
is identified at \LO{} with the quark radiated on the light-quark line and is referred to
as the spectator quark in the literature. This identification is only well-defined at \LO{}, of 
course.
The direction orthogonal to the plane made by the spectator quark
and the initial-state light-quark defines the $y$-axis, $\hat{y}$.
The initial-state light-quark direction, ($\vec{p}_q$), is chosen by
selecting the beam direction whose rapidity has
the same sign as that of the spectator jet.
Finally, the coordinate system is completed by defining $\hat{x}$ 
such that the system is right-handed \cite{Aguilar-Saavedra:2014eqa}.
Thus we have,
\begin{equation}
	\hat{z} = \frac{\vec{s}_t}{|\vec{s}_t|},\quad \hat{y} = \frac{\vec{s}_t \times 
		\vec{p}_q}{|\vec{s}_t \times 
		\vec{p}_q|},\quad \hat{x} = \hat{y} \times \hat{z} \label{eq:prodangles}\,.
\end{equation}
The (cosines of the) angles of the lepton with respect to these axes are referred to
as $\cosxprod, \cosyprod, \coszprod$.  Generically, these angles are
sensitive to operators that modify the production of the top quark.  

The second system uses as its first axis the direction of the $W$-boson in
the top-quark rest frame, $\hat{q}$. The second axis $\hat{N}$ is orthogonal to the
plane defined by $\hat{q}$ and the leading light-quark jet in the top-quark rest frame $\vec{s}_t$. 
As before, the system
is completed by the requirement of right-handedness, thus defining
$\hat{T}$~\cite{AguilarSaavedra:2010nx}:
\begin{equation} 
	\hat{q} = \frac{\vec{q}}{|\vec{q}|},\quad \hat{N} = \frac{ \vec{s}_t \times \vec{q} 
	}{|\vec{s}_t \times
		\vec{q}|},\quad \hat{T} = \hat{q} \times \hat{N} \label{eq:decayangles}\,.
\end{equation}
The angles between the lepton in the $W$-boson rest frame and these three
axes define the quantities $\coslstar$, $\coslN$ and $\coslT$.
Finally, we construct two more angles based on the projections of
the lepton in the $W$ boson rest frame onto the $\hat{N}$-$\hat{T}$
plane.  The angle between this projection and the $\hat{N}$ and $\hat{T}$ axes
define $\cosphiN$ and $\cosphiT$, respectively.
This second set of angles is particularly sensitive to modifications to
top-quark decay from physics beyond the Standard Model.

As has been noted before \cite{Neumann:2019kvk}, the
neutrino reconstruction has a noticeable impact on
most of these observables, since they are constructed in the top-quark rest frame which has
has a direct dependence on the neutrino four momentum. Here we do not consider
the neutrino reconstruction, but leave studying such effects for a future publication.

It was previously observed \cite{Sullivan:2005ar} that, after
cuts, going from \LO{} to \NLO{} had little effect on \SM{}
angular distributions like \coszprod{} that are used to measure $t$-channel single-top-quark
production, see also \cref{fig:coszprod}. We are now able to to quantify these findings at \NNLO{}
for the full set of observables defined above.

\begin{figure}
	\centering
	\includegraphics{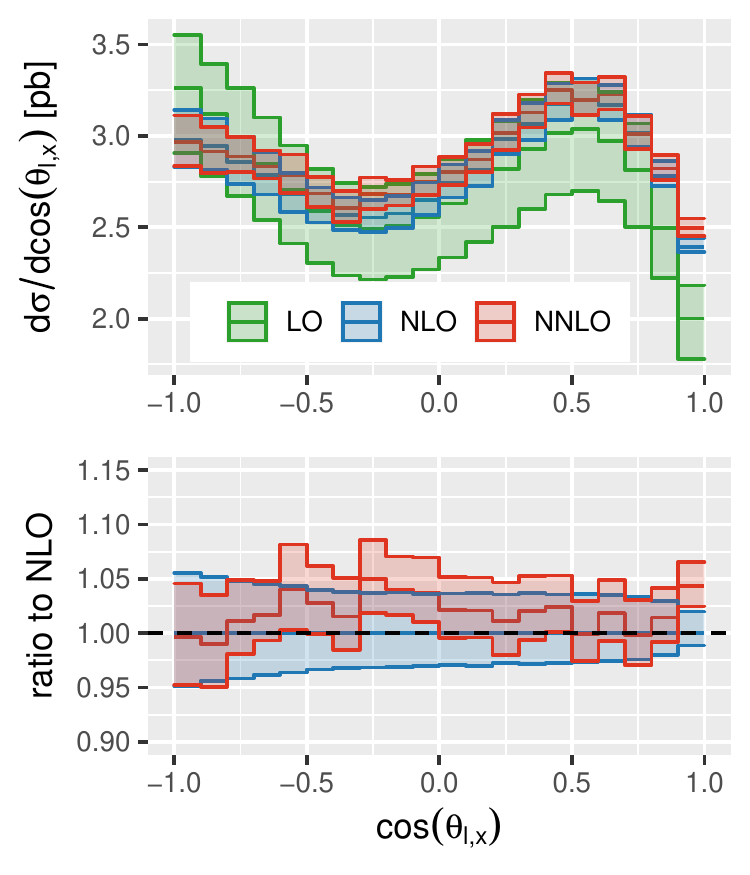}
	\includegraphics{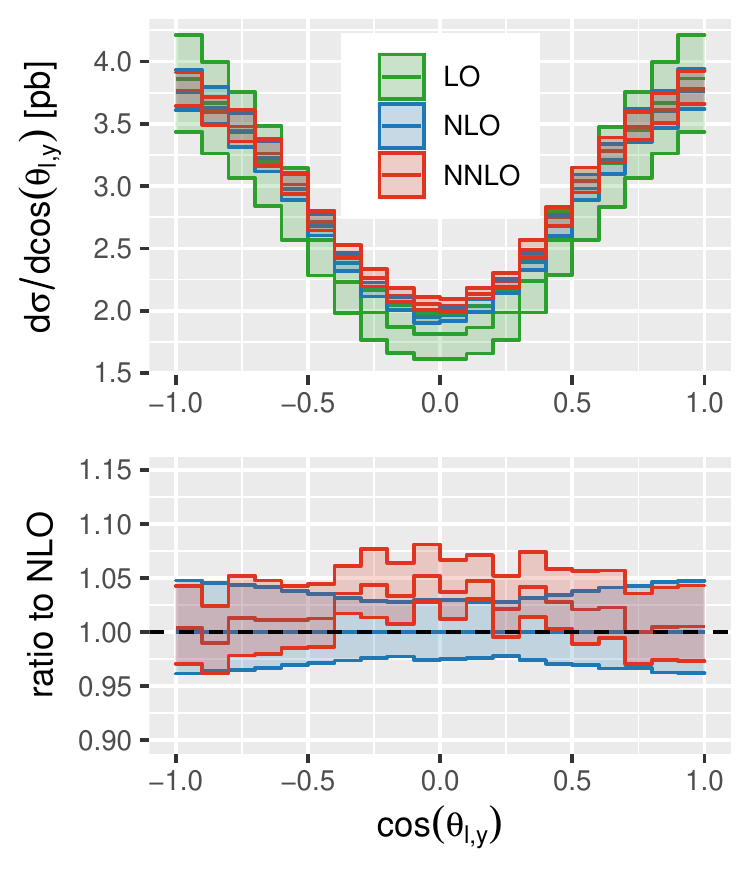}
	\caption{Angular distributions for $\cosxprod$ (left) and $\cosyprod$ (right).}
	\label{fig:cosxyprod}
\end{figure}

\begin{figure}
	\centering
	\includegraphics{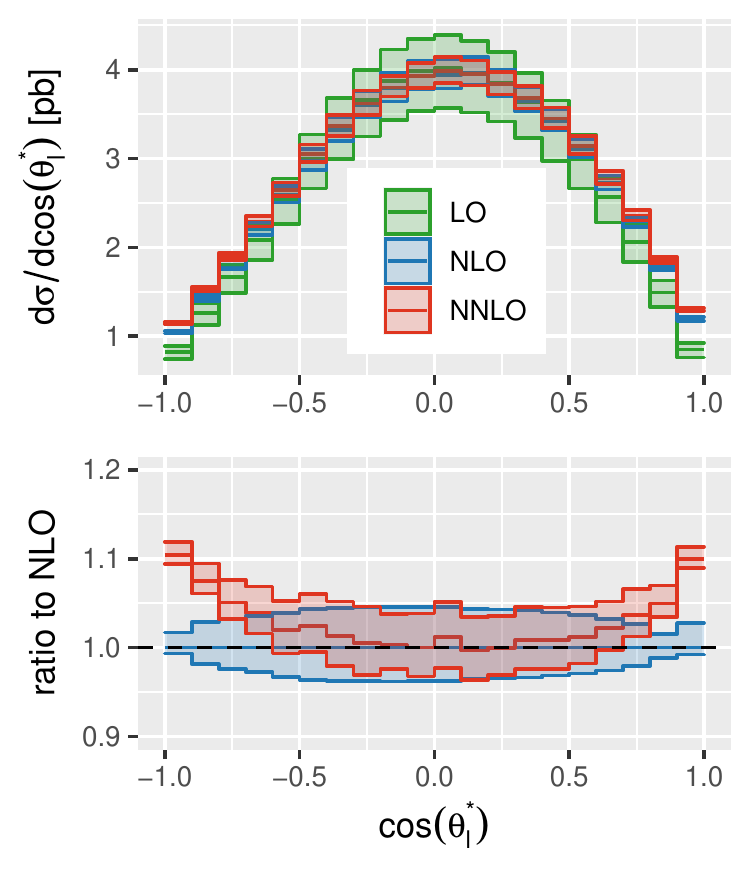}
	\caption{Angular distribution for $\coslstar$.}
	\label{fig:coslstar}
\end{figure}

\begin{figure}
	\centering
	\includegraphics{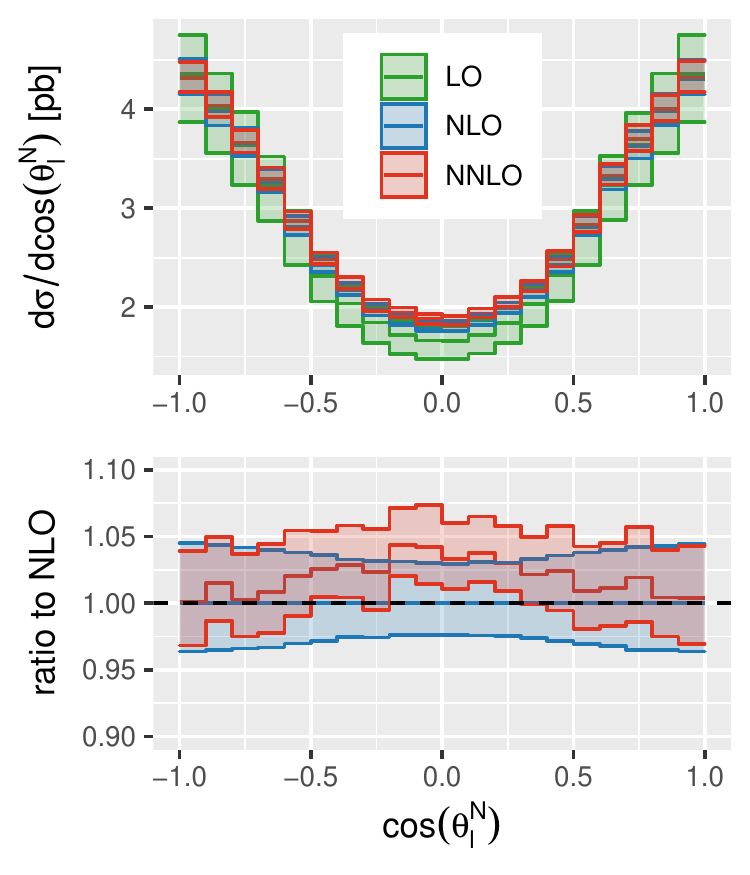}
	\includegraphics{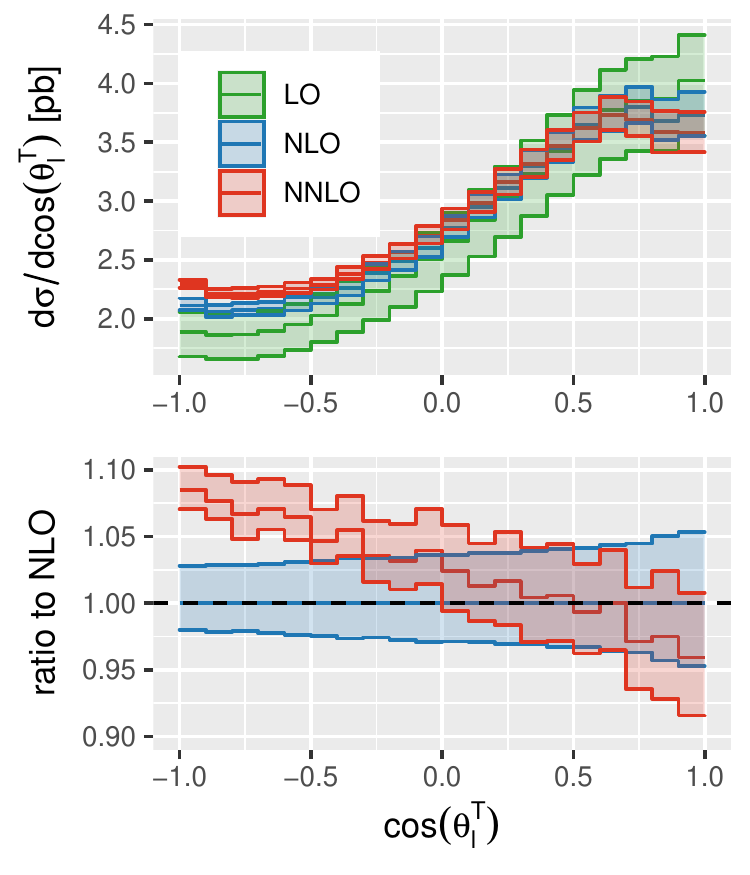}
	\caption{Angular distributions for $\coslN$ (left) and $\coslT$ (right).}
	\label{fig:coslNT}
\end{figure}

\begin{figure}
	\centering
	\includegraphics{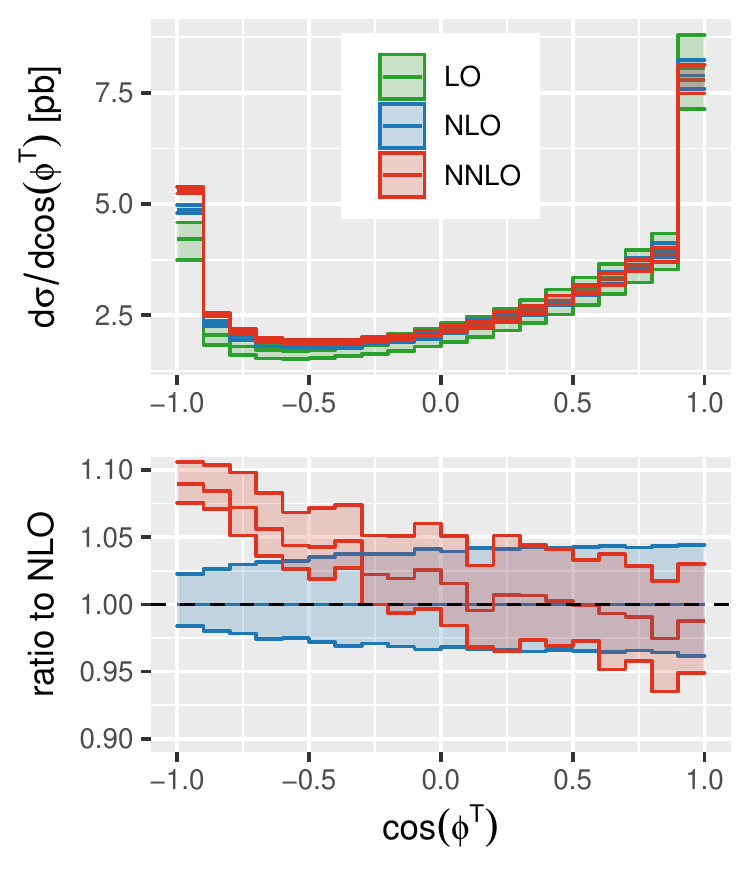}
	\includegraphics{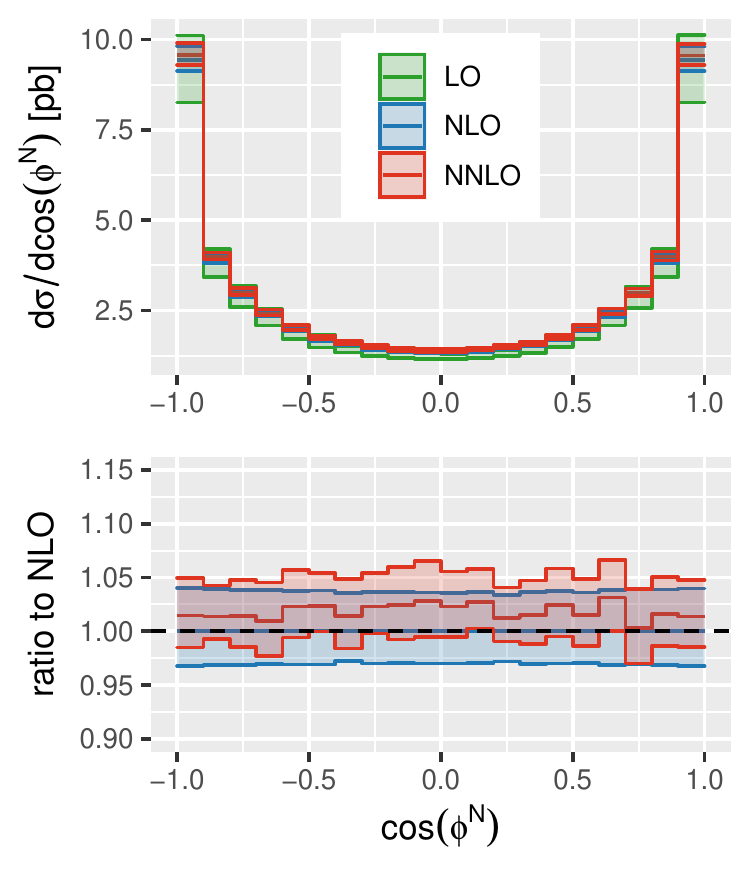}
	\caption{Angular distributions for $\cosphiT$ (left) and $\cosphiN$ (right).}
	\label{fig:cosphiNT}
\end{figure}

We present the results for \cosxprod, \cosyprod, \coszprod in \cref{fig:coszprod,fig:cosxyprod},
for \coslN, \coslT and \coslstar{} in \cref{fig:coslstar,fig:coslNT}, and \cosphiN and \cosphiT in 
\cref{fig:cosphiNT}. We find that all angular observables considered here are perturbatively 
stable and \NNLO{} corrections in the bulk are mostly consistent with zero within scale 
uncertainties.

Also off-shell effects have been considered before at 
\NLO{} \cite{Neumann:2019kvk} and were found to be relatively small and uniform at the few percent 
level, except for effects of up to $\sim 10\%$ in $\cosxprod$ and $\coslstar$. These effects
are therefore equally or more important than \NNLO{} on-shell corrections. Note that the 
$\coslN$ distribution has been found to be unstable in fixed-order perturbation theory due to
sensitivity to soft radiation \cite{Neumann:2019kvk}, but using the on-shell approximation we do not
expose this sensitivity.

\bibliographystyle{JHEP}
\bibliography{refs}

\end{document}